\documentclass[journal,10pt]{IEEEtran}

\usepackage{ifpdf}
\usepackage{cite}
\usepackage{url}

\usepackage{amsmath}
\usepackage{amssymb}
\usepackage{amsthm}
\usepackage{amsfonts}
\usepackage{multicol}
\usepackage{booktabs}
\usepackage{longtable}
\usepackage{bbding}
\usepackage{bm}
\usepackage{colortbl}
\usepackage[table]{xcolor}
\definecolor{LightCyan}{rgb}{0.88,1,1}
\definecolor{Gray}{gray}{0.9}
\usepackage{makecell}
\usepackage[pdftex]{graphicx}
\usepackage{graphicx}
\usepackage{epstopdf}
\usepackage{caption}
\usepackage{subfig}
\usepackage{float}
\usepackage{algorithm, algorithmic}
\usepackage{xcolor}
\newcommand\Mark[1]{\textsuperscript{#1}}   
\makeatletter
\def\ps@IEEEtitlepagestyle{
  \def\@oddfoot{\mycopyrightnotice}
  \def\@evenfoot{}
}
\def\mycopyrightnotice{
  {\footnotesize
  \begin{minipage}{\textwidth}
  \centering
  Copyright~\copyright~2015 IEEE. Personal use of this material is permitted. However, permission to use this  \\
  material for any other purposes must be obtained from the IEEE by sending a request to pubs-permissions@ieee.org.
  \end{minipage}
  }
}

\begin{document}

\title{User-Centric Cooperative Transmissions-enabled Handover for Ultra-Dense Networks}

\author{Nyaura~Mwinyi~Kibinda\Mark{a, b, c},~\IEEEmembership{Student Member,~IEEE},~Xiaohu~Ge\Mark{a, b},~\IEEEmembership{Senior Member,~IEEE}
\thanks{This work is supported by the National Natural Science Foundation of China under Grant U2001210. (\textit{Corresponding author: Xiaohu Ge.})}
\thanks{\textsuperscript{a}School of Electronic Information and Communications, Huazhong University of Science and Technology, Wuhan 430074, Hubei, Peoples's Republic of China (email: i201822022@hust.edu.cn; xhge@mail.hust.edu.cn).}
\thanks{\textsuperscript{b}China International Joint Research Center of Green Communications and Networking, Wuhan 430074, Peoples's Republic of China.}
\thanks{\textsuperscript{c}Department of Electronics and Telecommunications Engineering, The University of Dodoma, 41218 Dodoma, Tanzania.}}

\markboth{}
{}

\maketitle

\begin{abstract}
The user-centric cooperative transmission provides a compelling way to alleviate frequent handovers caused by an ever-increasing number of randomly deployed base stations (BSs) in ultra-dense networks (UDNs). This paper proposes a new user-centric cooperative transmissions-based handover scheme, i.e., the group-cell handover (GCHO) scheme, with the aim of reducing the handover rate in UDNs. In the proposed scheme, the boundary of the cooperating cluster depends on the distance among the user equipment (UE) and cooperating BSs. The new scheme captures the dynamicity and irregularity of the cooperating cluster topology resulting from randomly distributed BSs. Based on stochastic-geometry tools where BSs locations are modeled as the Poison point process (PPP), we derive an analytical expression of the handover rate for the UE with an arbitrary movement trajectory. Furthermore, a GCHO skipping (GCHO-S) scheme is proposed to minimize the handover cost, i.e., the percentage of time wasted in handover signaling in user-centric cooperative transmissions scenarios. The numerical results show that the GCHO scheme decreases the handover rate by 42.3\% and 72.7\% compared with the traditional single BS association and fixed-region cooperative network topology handover approaches, respectively. Moreover, under the same group-cell size and constant velocity, the GCHO-S scheme diminishes the handover cost by 50\% against the GCHO scheme.
\end{abstract}


\IEEEpeerreviewmaketitle

\section{INTRODUCTION}
\IEEEPARstart{T}{he} traffic growth of 175 zettabytes (175x$10^{21}$) is expected in the fifth generation (5G) cellular networks by 2025 \cite{b1}. The ultra densification deployment has been proposed as a breakthrough to meet the traffic requirements of 5G cellular networks. In 5G ultra-dense networks (UDNs), a large number of small cells with less service distance are installed inside a single macrocell area. Moreover, the massive multiple-input multiple-output (MIMO) and millimeter-wave (mmWave) transmission technologies motivate the deployment of 5G (UDNs) \cite{b2}, \cite{b3}. Small cells densification enhances the throughput and spectrum efficiency. However, substantial inter-cell interference and frequent handovers are emerging as great challenges in 5G UDNs \cite{b4,b5}.

In current 5G wireless networks, the user equipment (UE) handover from a source 5G base station (BS) to a target BS adopts the Xn interface, termed the Xn-based handover. The handover preparation and execution are performed over tunnels between the source and the target BS. The handover completion is carried out by the target BS and the core network through the N2 interface. The Xn-based handover procedure results in a few handover-related messages to the core network, subsequently lowering the signaling overhead \cite{b6}. However, the additional handover signaling overhead is inevitable in UDNs deployment due to frequent handover requests needed to be processed by the BSs (through the Xn interface) and the core network (through the N2 interface). In this paper, the handover from one BS to the other BS through the Xn interface is referred to as the traditional handover approach. Several studies have analyzed the impact of network densification on the traditional handover approach. For instance, the problem of handover latency in UDNs has been investigated in \cite{b7}. The results reveal that the frequent handovers caused by a rise in the densification ratio aggravate the handover latency. The authors in \cite{b8} corroborated that frequent mobility events in dense networks induce extra communication delay and lower the reliability performance. On a similar line, the evaluation model in \cite{b9} verified that the handover delay increases significantly for the fast-moving UE. Conclusively, for a specific radio link, frequent handovers increase the handover delay and signaling overhead in 5G UDNs based on the traditional handover method. Therefore, the frequent handover is a key issue impacting on the delay and signaling overhead of UDNs.

The user-centric cooperative transmission is recently deemed a distinctive technique to improve the handover performance of 5G UDNs. It involves a dynamic formation of serving multi-base station  cooperation in which the conventional BS-to-BS handover can be concealed \cite{b10}. In this regard, the handover enhancement approaches were studied in \cite{b11,b12,b13,b14}. A user-centric clustering-based handover management method was presented in \cite{b11}. The cooperation transmission gain was considered to facilitate seamless mobility while reducing the interruption ratio. In \cite{b12}, a received signal strength (RSS) prediction was used to select the target serving BSs for a multi-connectivity handover scheme to prevent frequent handovers. The proposed approach in \cite{b12} aimed to reduce over the air latency while simultaneously ensuring high reliability for packet delivery in ultra-reliable and low latency communication (URLLC). In \cite{b13}, the multi-BS cooperation mobility was characterized as the management of an anchor-based adaptive cooperative set to reduce the radio link failures in UDNs. A user-centric handover scheme was introduced in \cite{b14} to decrease the handover latency in ultra-dense low earth orbit (LEO) satellite networks. However, these studies focused on performance analysis without providing a tractable handover analytical framework, which could offer detailed insights for the handover performance improvement in UDNs. For example, it remains difficult to analyze the impact of the number of cooperating BSs, BS density and UE velocity on the handover performance.
{\tiny }
For accurate modeling and tractable analysis, the homogenous Poisson point process (PPP) and stochastic geometry tools \cite{b15,b16} have been commonly used to illustrate the handover management in dense and irregular deployed cellular networks \cite{b17, b18, b19, b20, b21, b22, b23, b24, b25}. The fundamental of mobility for single-tier cellular networks has been studied in \cite{b17}. Accounting for fractal characteristics in small cell networks, coverage probability and handover probability were obtained in \cite{b18}. In \cite{b19}, the handover probability was derived by employing the hysteresis margin in random UDNs. Yang  \textit{et al.} \cite{b20} analyzed the coverage and handover in dense mmWave networks with control and user plane (CUP) split architecture. On the basis of irregular cellular networks, the analytical frameworks for handover analysis in multi-tier heterogeneous cellular networks (HCN) have been studied in \cite{b21,b22,b23,b24}. In \cite{b25}, the coverage probability based on the distance between vehicles and cooperative BSs was presented in 5G networks considering co-channel interference. Moreover, the handover rate and vehicle overhead ratio were used to investigate the mobility performance of vehicles regarding cooperative transmissions.

Apart from that, several artificial intelligence (AI) techniques have been proposed in the literature to improve the handover performance in UDNs \cite{b26,b27,b28}. In \cite{b26}, the handover method based on multi-agent reinforcement learning was presented to reduce the frequent handovers while increasing the overall throughput. The handover and power allocation problem was modeled as a fully cooperative multi-agent to account for the interrelationship of different decisions made by various UEs. The deep Q learning technique based on the channel and load balancing characteristics of the UE has been exploited to optimize the handover procedure in UDNs \cite{b27}. With the purpose of enhancing mobility robustness while maintaining the quality of service (QoS) requirements, the Q learning method was studied in \cite{b28} to obtain an optimal handover triggering policy. It is worth noting that the previous research efforts only consider the handover analysis in the traditional handover scheme, i.e., BS-to-BS handover scheme. Thus, the presented methods are insufficient and cannot be extended to user-centric cooperative transmissions. Due to the spatial randomization of user-centric cooperating BSs in UDNs, an irregular cooperative cluster network topology is formed. Hence it is challenging to characterize the cluster boundary for the handover modeling in user-centric cooperative transmissions. Also, the handover in user-centric cooperative transmissions involves the dynamic change of serving BSs set \cite{b10}. At the same time, the dynamic change of BSs set may upsurge signaling overhead in the core network and undoubtedly hinder the gain provided by user-centric cooperative transmissions \cite{b29}, \cite{b30}. Accordingly, it is vital to characterize the user-centric cooperative handover with spatial randomness of BSs for 5G dense networks.

The research efforts on the handover design for user-centric cooperative transmissions have been presented in \cite{b31,b32,b33,b34}. The authors in \cite{b31} quantified the trade-off between the handover rate and data rate for multi-connectivity transmission to optimize the coordinated cluster size under non-coherent joint transmission (JT). The proposed model can be applied to the number-based and distance-based cooperation schemes. Considering the CUP separation architecture in multi-connectivity, the handover probability and handover cost-based average rate were derived in \cite{b32}. To analyze the effect of handover on coordinated multipoint (CoMP), the work in \cite{b33} derived the handover probability and coverage probability based on outdated channel state information (CSI) for 5G UDNs. The cell dwell time was explored in \cite{b34} to develop the movement-aware coordinated multipoint handover (MACH) scheme for the handover rate alleviation in heterogenous UDNs. Explicitly, the cell dwell time was used to formulate the movement trend of the mobile user. Based on the dwell time and the closest BS, an improved MACH (iMACH) scheme was proposed to improve the network reliability.

In an architectural context, the software-defined networking (SDN) has been introduced to lessen the BS multi-connectivity signaling overhead, which will undoubtedly be exacerbated with the network densification in the future 5G networks \cite{b35}, \cite{b36}. The SDN decouples the control and user planes hence improving the network’s manageability and adaptability. In \cite{b35}, \cite{b36}, the SDN controller was employed to conduct the BS association and cooperation. Through the SDN controller, the information exchange among cooperating BSs can be reduced significantly and the cooperation latency simultaneously decreases. As a consequence, the overall efficiency of the BSs cooperation can be improved.

However, the studies as mentioned above have the following limitations:
\begin{enumerate}
   \item The design perspective in \cite{b31} has ignored the dynamics of cooperative BSs cluster topology caused by random BSs placement. The authors in \cite{b31} considered the fixed-region group-cell topology for each UE, which results in similar handover numbers for all UEs in the network. As a matter of fact, each UE has a different interference environment in user-centric cooperating clusters. For this reason, the cluster network topology varies from each UE due to different distances between cooperating BSs and the UE, which leads to dissimilar statistics of handovers.
   \item Furthermore, the approaches in \cite{b32,b33,b34} have considered a standard path loss model in analyzing the impact of mobility on cooperative-based wireless networks, which is not realistic in dense wireless networks deployment \cite{b3}, \cite{b37}. Besides, circumventing frequent reformation of cooperating BSs regarding high dense BS deployment and/or fast-moving UEs remains an open problem in the current literature.
   \item Compared to the conventional cellular network, the UE needs to feedback the CSI to BSs for cooperating cluster formation in user-centric cooperative transmissions. In this case, the cooperation signaling overhead caused by the additional CSI feedback messages is proportional to the cluster size as indicated in \cite{b38,b39,b40}. However, the integrated impact of the number of cooperating BSs on the handover rate and cooperation signaling overhead has not been investigated for user-centric cooperative transmissions \cite{b31,b32,b33,b34}.
\end{enumerate}
To fully reap the benefits of network densification and user-centric cooperative transmissions, there is a need to address the aforementioned technical challenges judiciously and adequately.

This paper develops a handover scheme based on user-centric cooperative transmissions to reduce the number of handovers in UDNs. In the user-centric cooperation model, multiple BSs cooperation is orchestrated by the SDN controller. Considering BSs distributions as homogenous PPP, the closed-form expression of the handover rate is derived. Furthermore, this paper aims to give a first rigorous look at employing a dual-slope path loss model to analyze the performance of the mobile UE in user-centric cooperative networks. Motivated by studies in \cite{b41,b42,b43}, we propose a user-centric cooperation based handover skipping scheme to lessen the handover cost caused by the frequent cooperation cluster reformation in UDNs. Different from \cite{b41,b42}, our proposed handover skipping scheme involves dynamic user-centric cooperative transmissions. To the best of the author’s knowledge, this is the first time the handover skipping is explored in user-centric cooperative transmissions. Finally, the paper examines the benefit/cost tradeoff associated with the effect of the cooperating cluster size on the handover rate and cooperation signaling overhead. The main contributions of this paper are as follows
\begin{enumerate}
  \item We characterize the user-centric cooperative region for each UE as an interference protection region to offer a suitable balance between tractability and realism. The interference protection region is identified according to the distances among the UE and cooperating BSs. Based on the developed user-centric cooperative region, we propose a group-cell handover (GCHO) scheme and the handover rate is derived for a random movement trajectory.
  \item Based on the proposed GCHO scheme, the group-cell handover skipping scheme (GCHO-S) has been proposed to support the UE skipping the best-connected group-cell along the UE's trajectory for reducing frequent group-cell reformations.
  \item An analytical method for calculating the optimal group-cell size has been proposed to balance the tradeoff between the handover rate and cooperation signaling overhead in adopting the GCHO and GCHO-S schemes.
 \item Numerical results demonstrate that the proposed GCHO scheme decreases the handover rate against the traditional single BS association and fixed-region user-centric cooperative network topology handover approaches in UDNs. Moreover, the proposed GCHO-S scheme can effectively reduce the handover rate up to 50\% compared with the handover rate of the GCHO scheme in UDNs.
\end{enumerate}

The remainder of this paper is organized as follows. Section II introduces the system model. Section III describes the proposed group-cell handover scheme and derives a mathematical model for the handover rate. The coverage probability with the focus on the dual-slope path loss model is derived in Section IV. Moreover, Section IV analyzes the impact of handover rate on coverage probability and area spectral efficiency (ASE). The group-cell handover skipping scheme and the group-cell size optimization are presented in Section V. Numerical results are shown in Section VI. Finally, Section VII concludes the paper.
\section{SYSTEM MODEL}
This section presents the network model, channel propagation model and cooperation signaling overhead model. The commonly-used symbols and notations are summarized in Table I.
\subsection{Network Model}
\begin{figure}[!h]
\centering
\includegraphics[width=0.45\textwidth]{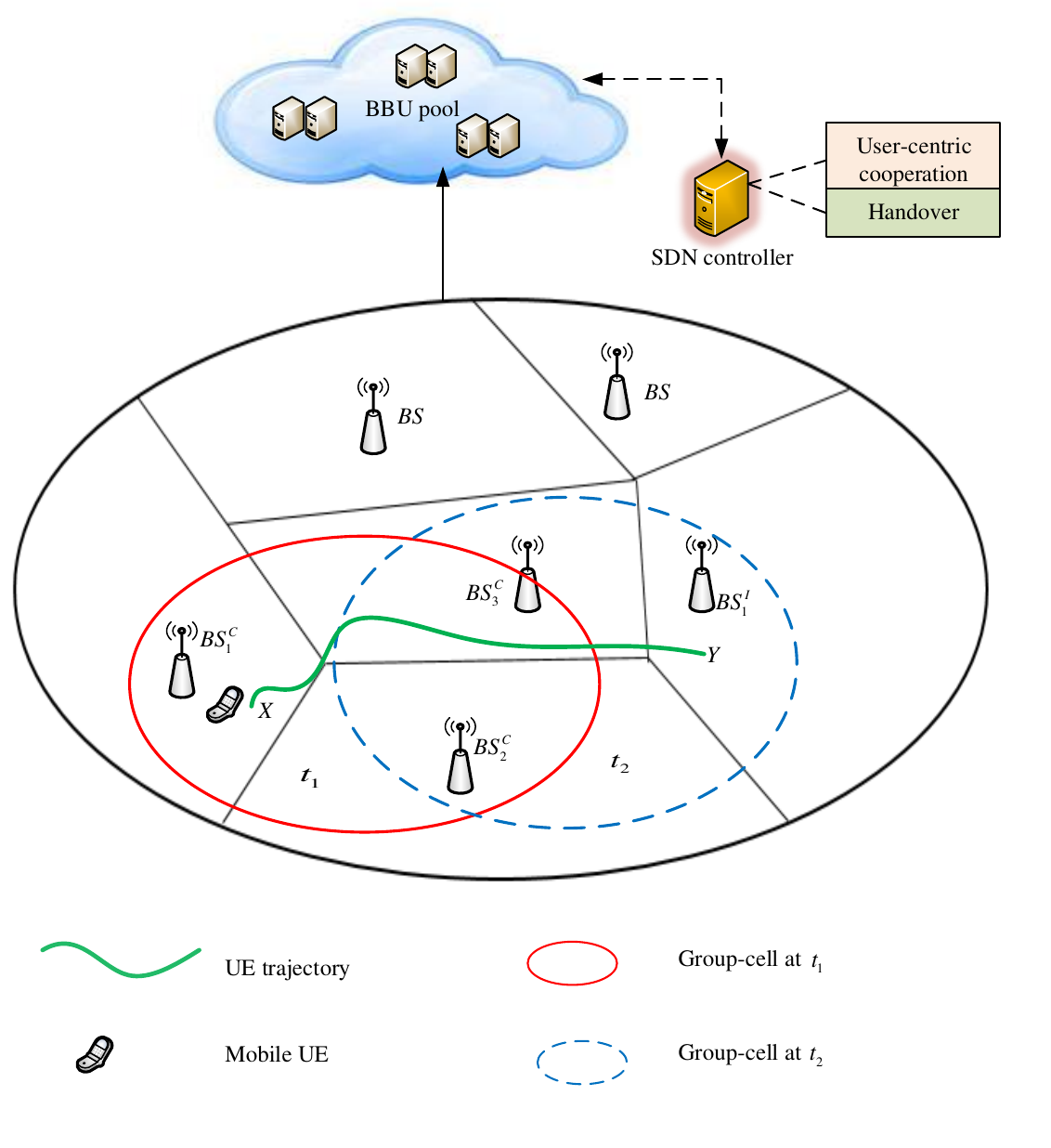}
\caption{Illustration of the group-cell transmission in SDN-based C-RAN architecture.}
\label{fig1}
\end{figure}

In this paper, the small cell UDNs are under-laid within the SDN based cloud-radio access network (C-RAN) architecture for user-centric cooperative transmissions in Fig. \ref{fig1}, \cite{b44}. The RAN functions are split between the baseband unit (BBU) hosted in the BBU pool and the remote radio units (RRUs) integrated into base stations (BSs).The SDN is implemented at the C-RAN level, where it decouples the control plane and data plane of the RAN for adaptive user-centric cooperation. In the control plane, the SDN controller intelligently handles the group-cell cooperation. Particularly, each BS transmits the CSI to the SDN controller, which returns the control information to the BSs for the group-cell formation. Furthermore, the SDN controller manages the group-cell handover\footnote{Throughout this work, “ the group-cell handover” refers to the scenario where a user dynamically changes the serving BSs set to support its movement, i.e., dynamic reformation of the group-cell.}. The BBU pool is connected to the SDN controller. The SDN data plane can provide the required data service for UEs. The OpenFlow protocol is employed to standardize the communication between the control plane and the data plane \cite{b36}.

In this paper, $G$ and $V$ are denoted as the sets of BSs and UEs, respectively. Let $g\in G$ and $v\in V$ denote the indexes of BSs and UEs, respectively. The locations of the BSs are deployed according to independent homogenous PPP ${{\Phi }_{BS}}$ with density ${{\lambda }_{BS}}$. Without loss of generality, the BSs are assumed to be deployed in an infinite two-dimension plane ${{\mathbb{R}}^{2}}$. To simplify the derivations, only the downlinks of small cell UDNs are considered. The cell coverages are split by the random Voronoi tessellation method  \cite{b35,b46}. Without loss of generality, a typical UE is picked randomly, and its initial location ${{Q}_{0}}\in {{\mathbb{R}}^{2}}$ is considered as the origin of the spatial coordinate. The UE at origin is referred to as a typical UE ${{U}_{v}}$. The UE  is associated with at least one BS. Each BS is indicated by $B{{S}_{g}}$ $\left( g=1,2,\,3,.... \right)$. In the group-cell scenario, the UE ${{U}_{v}}$ is jointly served by a set of cooperating BSs, denoted by $\mathbb{C}$ ($\mathbb{C}\subseteq G$). The group-cell that serves a typical UE is denoted by $K$. Also, in this paper, the group-cell size is marked as $M$, i.e., the number of cooperating BSs in the group-cell.
\subsection{Channel Propagation Model}
In UDNs, the BSs are positioned below the cluttered man-made structures, thus the radio signals in the line of sight (LoS) may encounter fewer absorption and diffraction losses than those in the non-line-of-sight (NLoS), leading to dissimilar path loss exponents. Hence, a widely used single-slope path loss model becomes inexact \cite{b37}, \cite{b47}. The path loss model with multiple slopes where different distance ranges are susceptible to different path loss exponents eventually is desirable in modeling UDNs scenarios [40]. In this work, the bounded dual-slope path loss model is expressed as follows \cite{b48}, \cite{b49}
\begin{equation} \label{eqn1}
\ell_{g}\left(\eta_{1}, \eta_{2} ; r_{g}\right)=\left\{\begin{array}{ll}
\left\|r_{g}\right\|^{-\eta_{1}}, & \left\|r_{g}\right\| \leq \mathcal{D}   \\
\Lambda\left\|r_{g}\right\|^{-\eta_{2}}, & \left\|r_{g}\right\|>\mathcal{D},
\end{array} \quad\right.
\end{equation}
where ${{\ell }_{g}}(.)$ denotes the path loss between the BS $B{{S}_{g}}$ and UE ${{U}_{v}}$. $\left\| {{r}_{g}} \right\|$ is the Euclidian distance between the UE ${{U}_{v}}$  and BS $B{{S}_{g}}$. $\mathcal{D}$ ($\mathcal{D}>0$) is the critical distance that is used to approximate the LoS $\left( \left\| {{r}_{g}} \right\|\le D \right)$ and NLoS $\left(\left\| {{r}_{g}} \right\|>\mathcal{D} \right)$ regimes. ${{\eta }_{1}}$ and ${{\eta }_{2}}$ are the path loss exponents for LoS and NLoS propagation signals, respectively with $0\le {{\eta }_{1}}\le {{\eta }_{2}}$. $\Lambda$ is introduced to maintain continuity and is constant,  $\Lambda \triangleq \mathcal{D}^{{{\eta }_{2}}-{{\eta }_{1}}}$.

The BSs and UEs are assumed to be equipped with omnidirectional antennas\footnote{Note that the proposed analytical model can be readily extended to massive MIMO scenario, but its impact on network performance requires further exploration.}. Perfect CSI is assumed to be obtained at the transmitter and receiver, respectively. The UE ${{U}_{v}}$ associates with the BS which provides the minimum path loss \cite{b50}. We consider an interference-limited scenario in urban environment. Therefore, the thermal noise is neglected. Accordingly, based on the user-centric cooperative transmissions, the signal-to-interference ratio (SIR) experienced by the UE ${{U}_{v}}$ located at the origin is calculated as
\begin{equation}\label{eqn2}
\begin{aligned}
  & SIR(M)=\frac{\sum\limits_{g=1}^{M}{{{p}_{g}}{{h}_{g}}\times 1\left( {{\ell }_{g}}({{\eta }_{1}},{{\eta }_{2}};{{r}_{g}})\ge \beta  \right)}}{\sum\limits_{g>M}{{{p}_{g}}\Lambda r_{g}^{-{{\eta }_{2}}}{{h}_{g}}}}, \\
\end{aligned}
\end{equation}
where ${{p}_{g}}$ is the transmission power in each BS $B{{S}_{g}}$ and ${{h}_{g}}$ is the channel power gain between the BS $B{{S}_{g}}$ and UE ${{U}_{v}}$ under Rayleigh fading. Without loss of generality, ${{p}_{g}}$ is normalized to 1 and a random variable ${{h}_{g}}$ follows an exponential distribution with mean 1 and denoted as ${{h}_{g}}\sim \exp (1)$. Furthermore, $\sum\limits_{g>M}{{{p}_{g}}\Lambda r_{g}^{-{{\eta }_{2}}}{{h}_{g}}}$ indicates the aggregate co-channel interference created by non-group-cell BSs, $\beta $ is the group-cell cooperation threshold and $\sum\limits_{g=1}^{M}{{{p}_{g}}{{h}_{g}}\times 1\left( {{\ell }_{g}}({{\eta }_{1}},{{\eta }_{2}};{{r}_{g}})\ge \beta  \right)}$ is the received signal power from cooperating BSs.

\subsection{Cooperation Signaling Overhead Model}
In this paper, the cooperation signaling overhead is modeled as a function of the group-cell size. In user-centric cooperative transmissions, the UE ${{U}_{v}}$ estimates the CSI and provides the CSI feedback to the cooperating BSs. Then, the cooperating BSs forward the received CSI to the SDN controller for adaptive cooperation \cite{b38}, \cite{b39}. From this perspective, we define the cooperation signaling overhead as the total number of the CSI messages fed back by the UE ${{U}_{v}}$ to the SDN controller through cooperating BSs in a time interval $T$. Without loss of generality, $T$ is assumed to be identical for all cooperating BSs in the group-cell $K$. A linear correspondence between the number of the CSI feedback messages and group-cell size is presented in the literature \cite{b40}. Hence, the accumulated cooperation signaling overhead for the group-cell $K$ with size $M$ is expressed as
\begin{equation}\label{eqn3}
   \Upsilon _{K}=\frac{\mu }{T}M,
\end{equation}
where $\mu$ is the number of feedback messages from the UE ${{U}_{v}}$ to the BS $B{{S}_{g}}$. For simplification, we assume that $\mu $ is the same for each BS $B{{S}_{g}}$. In this work, we focus only on the feedback signaling overhead resulting from user-centric cooperative transmissions. Assumed that the CSI feedback links have unlimited capacity, the CSI feedback messages are transmitted successfully without errors.
\begin{table}[!t]
\caption{ LIST OF NOTATIONS}
\label{table 1}
\centering
\begin{tabular}{|c||c|}
\hline
  Symbol&Description\\
\hline
 $SIR(M)$& SIR in the group-cell \\
 $p^{M}(\tau ,{{\lambda }_{BS}})$&	Group-cell coverage probability   \\
  $M$&	Group-cell size  \\
  $\mathcal{D}$&	Critical distance  \\
  ${{\mathcal{L}}_{I(M)}}(s)$&	Laplace transform   \\
  $\tau $&	SIR threshold   \\
  ${{\eta }_{1}}$, ${{\eta }_{2}}$& Path loss exponents   \\
  ${{H}_{M}}$&	Group-cell handover rate   \\
  ${{\mathfrak{T}}_{v}}$&	Trajectory length \\
  ${{\Theta }_{{{r}_{M}}}}$& Set of group-cell boundaries\\
  ${{\Omega }_{1}}({{\Theta }_{{{r}_{M}}}})$	& Length intensity of the group-cell boundary \\
  ${{\delta }_{M}}$& Extended group-cell boundary \\
  ${{\Omega }_{2}}({{\Theta }_{{{\delta }_{M}}}})$&	Area intensity of the extended group-cell boundary \\
  $\Psi $&	UE Velocity \\
  $H_{M}^{*}$& Handover rate during GCHO skipping \\
  $r_{1}^{I}$& First nearest distance outside the group-cell \\
  $r_{2}^{I}$& Second nearest distance outside the group-cell \\
  $\Upsilon _{K}$ &	Cooperation signaling overhead \\

\hline
\end{tabular}
\end{table}

\section{GROUP-CELL HANDOVER RATE MODEL}
Considering the randomness of BSs locations and the dynamic group-cell network topology, the group-cell and the group-cell handover are characterized and modeled in this section.

\begin{figure}[!h]
\centering
\includegraphics[width=0.45\textwidth]{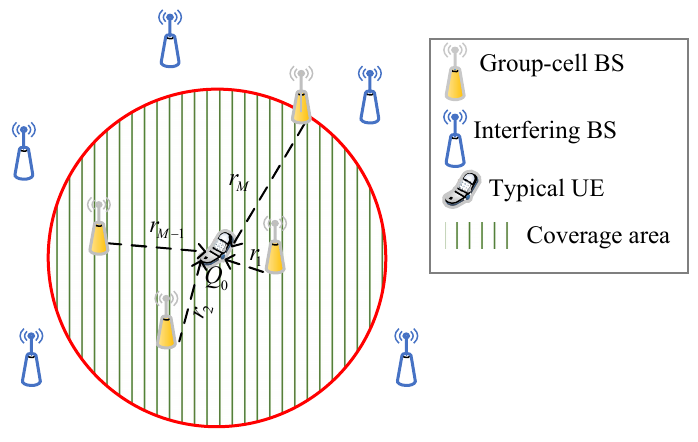}
\caption{Group-cell abstraction from user-centric cooperation in UDNs for a typical UE. }
\label{fig2}
\end{figure}

  We assume that the UE ${{U}_{v}}$ is initially located at the origin with the association distance ${{r}_{1}},\,\,{{r}_{2}},.....{{r}_{M}}$ between the UE ${{U}_{v}}$ and the BSs within the group-cell. Since the locations of the BSs in UDNs are random, the distance between the UE ${{U}_{v}}$ and cooperating BSs are generally randomly. To simplify calculations, the distance of serving BSs are configured to be ascendingly ordered from the mobile UE's initial location, which can be expressed as  ${{r}_{1}}<{{r}_{2}}.....<{{r}_{M}}$. The group-cell region is portrayed as a disk region around the UE ${{U}_{v}}$. We assume that there is no interference among BSs in the group-cell region. In this paper, the group-cell region is termed as an interference protection region with the tuple $\left\{ {{r}_{M}},{{A}^{M}} \right\}$, as illustrated in Fig. \ref{fig2}. ${{r}_{M}}$ is the radius of the interference protection region for the $M-th$ BS in the group-cell of size $M$. ${{A}^{M}}$ denotes the geographic coverage area of the group-cell. The distance ${{r}_{M}}$ is measured during the group-cell configuration and updated after each successful handover according to UE's measurement report. Considering ${{r}_{M}}$ is a random variable, then the distance distribution function of ${{r}_{M}}$ in ${{\mathbb{R}}^{2}}$ can be expressed as \cite{b51}
\begin{equation}\label{eqn4}
{{f}_{{{r}_{M}}}}(M)=\frac{2{{\left( \pi {{\lambda }_{BS}} \right)}^{M}}}{\Gamma (M)}{{e}^{-{{\lambda }_{BS}}\pi {{r}_{M}}}}^{2}{{r}_{M}}^{2M-1} ,
\end{equation}
where $\Gamma (.)$ is the Gamma function for a positive integer $M$, $\Gamma (M)=\left( M-1 \right)!$.

As it is shown in Fig. \ref{fig2}, the whole region can be divided into two areas. The BSs inside the disk are the cooperating BSs, i.e., interference protection region BSs and the BSs outside the disk are the interfering BSs. Let $BS_{j}^{C}$, $j\in \left( 1,\,\,2,\,.\,.\,..M-1,\,M \right)$, denotes the cooperating BS in the group cell. In this case, the $BS_{M}^{C}$ is the farthest associated BS in the group-cell. Let $BS_{q}^{I}$, $q=(1,\,2,\,3.....)$, denotes the interfering BS. When $q=1$, i.e., $BS_{1}^{I}$ is the first nearest interfering BS corresponding to the $BS_{M}^{C}$. We assume that the UE ${{U}_{v}}$ moves from its original location towards the farthest associated BS $BS_{M}^{C}$ on an arbitrary long trajectory with the velocity $\Psi$. The moving direction of the UE ${{U}_{v}}$ is assumed to be uniformly distributed in $[0,2\pi ]$ \cite{b52}. The serving group-cell is reformed dynamically to follow the serviced UE mobility. The group-cell handover (GCHO) scheme is triggered when the UE ${{U}_{v}}$ crosses the group-cell boundaries. As a result, the set of connected BSs is modified to adapt to the refreshed network topology. The associated BS $BS_{1}^{C}$ in the group-cell, which is obviously near to the UE ${{U}_{v}}$ before movement, is assumed to become the farthest BS when the UE ${{U}_{v}}$ crosses the group-cell boundaries. The SDN controller will determine whether adds a new BS $BS_{1}^{I}$ to the group-cell while removing the BS $BS_{1}^{C}$ in the group-cell concurrently. This paper assumes that only one of the associated BSs is altered. Therefore, there are $M-1$ common BSs in the newly formed group-cell after every GCHO. Fig. \ref{fig1} exemplifies the GCHO for the given UE trajectory, with $M=3$ . The  UE ${{U}_{v}}$ moves from point X to point Y with the corresponding set of cooperative BSs, ${{\mathbb{C}}_{1}}=\left\{ BS_{1}^{C},\,BS_{2}^{C},\,BS_{3}^{C} \right\}$ and ${{\mathbb{C}}_{2}}=\left\{ BS_{2}^{C},\,BS_{3}^{C},\,BS_{1}^{I} \right\}$ at timeslot ${{t}_{1}}$ (before GCHO) and timeslot ${{t}_{2}}$ (after GCHO), respectively.

To model the handover rate in user-centric cooperative transmissions, the characterized group-cell region is used to track the number of handovers. We assume that the UE ${{U}_{v}}$  moves from its original location to the farthest associated BS $B{{S}_{M}}$ with a trajectory ${{\mathfrak{T}}_{v}}$ of finite length $\left| {{\mathfrak{T}}_{v}} \right|$. The GCHO is executed when the UE ${{U}_{v}}$  crosses the boundaries of the group-cell region. We denote ${{\Theta }_{{{r}_{M}}}}$ as the set of the group-cell boundaries in ${{r}_{M}}$. The expected handovers number of the UE ${{U}_{v}}$  is represented by $\mathbb{E}\left( N({{\mathfrak{T}}_{\nu }},{{\Theta }_{{{r}_{M}}}}) \right)$, where $\mathbb{E}\left( . \right)$ denotes the expectation operation. The corresponding group-cell handover rate ${{H}_{M}}$  in a time $\Delta t$ is calculated by,
\begin{equation}\label{eqn5}
  {{H}_{M}}=\frac{\mathbb{E}\left( N({{\mathfrak{T}}_{v}},{{\Theta }_{{{r}_{M}}}}) \right)}{\mathbb{E}(\Delta t)},
\end{equation}

For tractability, we use the generalized argument of Buffon's needle problem to derive the expected number of handovers during one movement period of the UE ${{U}_{v}}$ \cite{b53}, \cite{b17}. The group-cell boundaries are described by the length intensity ${{\Omega }_{1}}({{\Theta }_{{{r}_{M}}}})$, which is characterized as the average length of ${{\Theta }_{{{r}_{M}}}}$ in a unit square from which we can track the number of intersections $N({{\mathfrak{T}}_{v}},{{\Theta }_{{{r}_{M}}}})$ \cite{b31}, \cite{b45}. Since ${{\Theta }_{{{r}_{M}}}}$ is stationary and isotropic, the picking of the unit square can be randomly on ${{\mathbb{R}}^{2}}$. Conditioned on the velocity and the direction of the UE, the group-cell handover rate for a typical UE ${{U}_{v}}$  located in an area $\left| {{A}^{M}} \right|$ is derived as [45, Secs. 8.3, 9.3 and 10.6 ]
\begin{equation}\label{eqn6}
  \begin{aligned}
  & {{H}_{M}}=\frac{2}{\pi }\Psi \underset{\left| {{A}^{M}} \right|\to \infty }{\mathop{\lim }}\,\frac{\left| {{\Theta }_{{{r}_{M}}}} \right|}{\left| {{A}^{M}} \right|} \\
 & \,\,\,\,\,\,\,\,\,\,=\frac{2}{\pi }\Psi {{\Omega }_{1}}({{\Theta }_{{{r}_{M}}}})\,. \\
\end{aligned}
\end{equation}
With
\begin{equation}\label{eqn7}
{{\Omega }_{1}}({{\Theta }_{{{r}_{M}}}})=\underset{{{\delta }_{M}}\to 0}{\mathop{\lim }}\,\frac{{{\Omega }_{2}}({{\Theta }_{{{\delta }_{M}}}})}{2{{\delta }_{M}}}\,,
\end{equation}
where ${{\delta }_{M}}$ denotes the extended group-cell boundary, ${{\Theta }_{{{\delta }_{M}}}}$ is set of the extended group-cell boundaries due to ${{\delta }_{M}}$, and ${{\Omega }_{2}}({{\Theta }_{{{\delta }_{M}}}})$ is the area intensity. The area intensity is defined as the expected area of ${{\Theta }_{{{\delta }_{M}}}}$ in a unit square. Also, the point is considered in the ${{\Theta }_{{{\delta }_{M}}}}$ provided that its shortest perpendicular distance to ${{\Theta }_{{{r}_{M}}}}$ is less than ${{\delta }_{M}}$.\\
\textbf{Theorem 1:} Given a ${{\Theta }_{{{\delta }_{M}}}}=\left\{ {{\varpi }_{2}}\in {{\mathbb{R}}^{2}}|\exists {{\varpi }_{1}}\,\,\in {{\Theta }_{{{r}_{M}}}},\,\,s.t.\,\,\left| {{\varpi }_{2}}-{{\varpi }_{1}} \right|<{{\delta }_{M}} \right\}$, where ${{\varpi }_{1}}$ is the point in the group-cell boundary and ${{\varpi }_{2}}$ is any other point in ${{\mathbb{R}}^{2}}$, then the area intensity of ${{\Theta }_{{{\delta }_{M}}}}$ is
\begin{equation}\label{eqn8}
  {{\Omega }_{2}}({{\Theta }_{{{\delta }_{M}}}})=\frac{2{{\delta }_{M}}}{{{r}_{M}}}+O({{\delta }_{M}}^{2})\,,
\end{equation}
where $O(.)$ denotes the second-order term of ${{\delta }_{M}}$. \\
\textbf{Proof:} Please refer to Appendix A.

By substituting (\ref{eqn8}) into (\ref{eqn7}), the length intensity can be expressed by
\begin{equation}\label{eqn9}
  {{\Omega }_{1}}({{\Theta }_{{{r}_{M}}}})=\frac{1}{{{r}_{M}}},
\end{equation}
and by using (\ref{eqn6}), we can derive the handover rate as
\begin{equation}\label{eqn10}
  {{H}_{M}}=\frac{2}{\pi {{r}_{M}}}\Psi.
\end{equation}
According to (\ref{eqn9}), the length intensity ${{\Omega }_{1}}({{\Theta }_{{{r}_{M}}}})$ decreases with the increase of the interference protection region radius ${{r}_{M}}$. Since the BSs are deployed as homogenous PPP, the coverage area $\left| {{A}^{M}} \right|$ is bounded by $\frac{M}{{{\lambda }_{BS}}}$ \cite{b54}. With this setting, the handover rate in (\ref{eqn10}) can be rewritten as
\begin{equation}\label{eqn11}
  {{H}_{M}}=\frac{2\Psi \sqrt{{{\lambda }_{BS}}}}{\sqrt{\pi }\sqrt{M}}.
\end{equation}
\section{COVERAGE PROBABILITY WITH GROUP-CELL HANDOVER SCHEME}
In this section, the coverage probability is derived based on the dual-slope path loss model to study the impact of handover cost on user-centric cooperative transmissions.
\subsection{Coverage Probability}
The coverage probability is stated as the probability that the received SIR in the group-cell is larger than the threshold $\tau $ and is given by
\begin{equation}\label{eqn12}
 p^{M}(\tau ,{{\lambda }_{BS}})\triangleq \mathbb{P}(SIR(M)>\tau ).
\end{equation}
Considering the dual-slope path loss model, the $SIR(M)$ is expressed as
\begin{equation}\label{eqn13}
SIR(M)=\frac{\sum\limits_{g=1}^{J}{{{p}_{g}}r_{g}^{-{{\eta }_{1}}}h_{g}^{*}}+\sum\limits_{g=J+1}^{M}{{{p}_{g}}\Lambda r_{g}^{-{{\eta }_{2}}}h_{g}^{**}}}{I(M)},
\end{equation}
where $h_{g}^{*}$ and $h_{g}^{**}$ are the channel power gain between the BS $B{{S}_{g}}$ and UE ${{U}_{v}}$ for ${{r}_{g}}\le {\mathcal{D}}$ (LoS) and ${{r}_{g}}>{\mathcal{D}}$ (NLoS), respectively. The aggregate interference of non-cooperated BSs is denoted by $I(M)=\sum\limits_{g>M}{{{p}_{g}}\Lambda r_{g}^{-{{\eta }_{2}}}{{h}_{g}}}$. By substituting (\ref{eqn13}) into (\ref{eqn12}), the coverage probability of the typical UE ${{U}_{v}}$ located at the origin is expressed as
\begin{equation} \label{eqn14}
\begin{aligned}
  & {{p}^{M}}(\tau ,{{\lambda }_{BS}})\triangleq \mathbb{P}(SIR(M)>\tau ) \\
 & \,\,\,=\mathbb{P}\left( \frac{\sum\limits_{g=1}^{J}{{{p}_{g}}r_{g}^{-{{\eta }_{1}}}h_{g}^{*}}+\sum\limits_{g=J+1}^{M}{{{p}_{g}}\Lambda r_{g}^{-{{\eta }_{2}}}h_{g}^{**}}}{I(M)}>\tau  \right)\, \\
 & \,\,={{\mathbb{E}}_{{{r}_{g}}}}\left( \mathbb{P}\left( \begin{aligned}
  & \sum\limits_{g=1}^{J}{{{p}_{g}}r_{g}^{-{{\eta }_{1}}}h_{g}^{*}} \\
 & +\sum\limits_{g=J+1}^{M}{{{p}_{g}}\Lambda r_{g}^{-{{\eta }_{2}}}h_{g}^{**}}>\tau I(M) \\
\end{aligned} \right) \right) \\
\end{aligned},
\end{equation}
where $\sum\limits_{g=1}^{J}{h_{g}^{*}}$ and $\sum\limits_{g=J+1}^{M}{h_{g}^{**}}$ are independent Gamma distributed random variables with $\sum\limits_{g=1}^{J}{h_{g}^{*}}\sim \Gamma \left( J,1 \right)$ and $\sum\limits_{g=J+1}^{M}{h_{g}^{**}}\sim \Gamma \left( M-J,1 \right)$, respectively. Subsequently, the summation of $\sum\limits_{g=1}^{J}{r_{g}^{-{{\eta }_{1}}}h_{g}^{*}}\sim \Gamma \left( J,\frac{1}{\sum\limits_{g=1}^{J}{r_{g}^{-{{\eta }_{1}}}}} \right)$and $\sum\limits_{g=J+1}^{M}{\Lambda r_{g}^{-{{\eta }_{2}}}h_{g}^{**}}\sim \Gamma \left( M-J,\frac{1}{\sum\limits_{g=J+1}^{M}{\Lambda r_{g}^{-{{\eta }_{2}}}}} \right)$ are the convolutions\footnote{Let  $\sum\limits_{g=1}^{J}{r_{g}^{-{{\eta }_{1}}}h_{g}^{*}}=A$,$\sum\limits_{g=J+1}^{M}{\Lambda r_{g}^{-{{\eta }_{2}}}h_{g}^{**}}=B $ and $z=\tau I(M)$, then $\mathbb{P}\left( A+B>z \right)=1-\left( {{F}_{A}}\left( z \right)\,\otimes{{F}_{B}}\left( z \right) \right)=1-\int\limits_{0}^{z}{{{F}_{A}}\left( z-b \right){{f}_{B}}(b)d}b\,\,\,$, where $(\otimes )$  denotes the convolution.} of each of the corresponding distributions.

Equation (\ref{eqn14}) is difficult to work with directly due to the inaccessible probability distribution function of the random variable ${{r}_{g}}$ in both LoS and NLoS regimes. To get a better handle on $SIR(M)$, an approximation form of coverage probability ${{p}^{M}}(\tau ,{{\lambda }_{BS}})$  is of importance. Since the service distance ${{r}_{g}}$ of BSs is limited in a cooperative group-cell region considering UDNs, the wireless links among cooperative BSs are assumed as LoS links in this paper. Moreover, the interfering links outside the group-cell region are regarded as NLoS\footnote{Note that, such assumption is mainly done for tractability. Its accuracy was validated in \cite{b59} by the simulation results and has provided the useful performance and observations. Hence, the validation is omitted for brevity in this paper.} links. Furthermore, the predefined cooperation threshold $\beta $ is viewed as a function of the critical distance ${\mathcal{D}}$. Thus, the $SIR(M)$ can be rewritten as
\begin{equation}\label{eqn15}
  SIR(M)=\frac{\sum\limits_{g=1}^{M}{{{p}_{g}}r_{g}^{-{{\eta }_{1}}}{{h}_{g}}}}{I(M)}.
\end{equation}
The coverage probability ${{p}^{M}}(\tau ,{{\lambda }_{BS}})$ is further derived in Theorem 2.\\
\textbf{Theorem 2:} Under the condition of having the group-cell BSs within the LoS link, the coverage probability of the UE ${{U}_{v}}$ located at the edge of cooperating BSs is expressed by (\ref{eqn16}), where $R$ is the distance between a cell edge UE and cooperating BSs. \\
\begin{figure*}
\begin{equation}\label{eqn16}
p^{M}(\tau ,{{\lambda }_{BS}})=\int\limits_{0}^{\infty }{2{{\left( \pi {{\lambda }_{BS}} \right)}^{2}}{{R}^{3}}{{e}^{-\pi {{\lambda }_{BS}}{{R}^{2}}}}\left( \exp \left( -\pi {{\lambda }_{BS}}{{\left( \tau \mathcal{D}^{{{\eta }_{2}}-{{\eta }_{1}}}{{R}^{-{{\eta }_{1}}}} \right)}^{\frac{2}{{{\eta }_{2}}}}}\int\limits_{\vartheta }^{\infty }{\left( \frac{1}{1+{{u}^{\tfrac{{{\eta }_{2}}}{2}}}} \right)du} \right)+{{A}_{M-1}} \right)dR}.
\end{equation}
\end{figure*}
\textbf{Proof:} Please refer to Appendix B.

To evaluate the handover effect on the coverage probability, the handover cost-aware coverage probability bears the following expression \cite{b21}, \cite{b55}: \\
\begin{equation}\label{eqn17}
{{p}^{\tilde M}}={{p}^{M}}(\tau ,{{\lambda }_{BS}})\times {{e}_{h}}(1-{{d}_{\cos t}})+{{p}^{M}}(\tau ,{{\lambda }_{BS}})\times \left( 1-{{e}_{h}} \right),
\end{equation}
with
\begin{equation}\label{eqn18}
  {{d}_{cost}}={{t}_{H}}{{H}_{M}},
\end{equation}
where ${{d}_{cost}}$ is the handover cost and defined as the percentage of time wasted in handover signaling, i.e., the service delay. It is also assumed that no data is transmitted during the handover processes. The delay incurred by each handover is denoted as ${{t}_{H}}$. The handover event indicator is ${{e}_{h}}$ (${{e}_{h}}\in (0,1)$), where ${{e}_{h}}=0$ and ${{e}_{h}}=1$ are corresponding to no handover and handover events, respectively. Additionally, based on the results in \cite{b56}, the handover cost-aware ASE considering user mobility can be written as
\begin{equation}\label{eqn19}
ASE_{cost}^{M}={{\lambda }_{BS}}{{\log }_{2}}(1+\tau )\times {{p}^{\tilde M}}.
\end{equation}

\subsection{Simulation Analysis}
In this subsection, the numerical results of the coverage probability performance are presented for three cooperating BSs, i.e., $M$=3. Explicitly, the default parameters are configured as follows: the path loss exponents are ${{\eta }_{1}}=2$ and ${{\eta }_{2}}=4$ unless noted otherwise.\\
\begin{figure}[!h]
\centering
\includegraphics[width=0.45\textwidth]{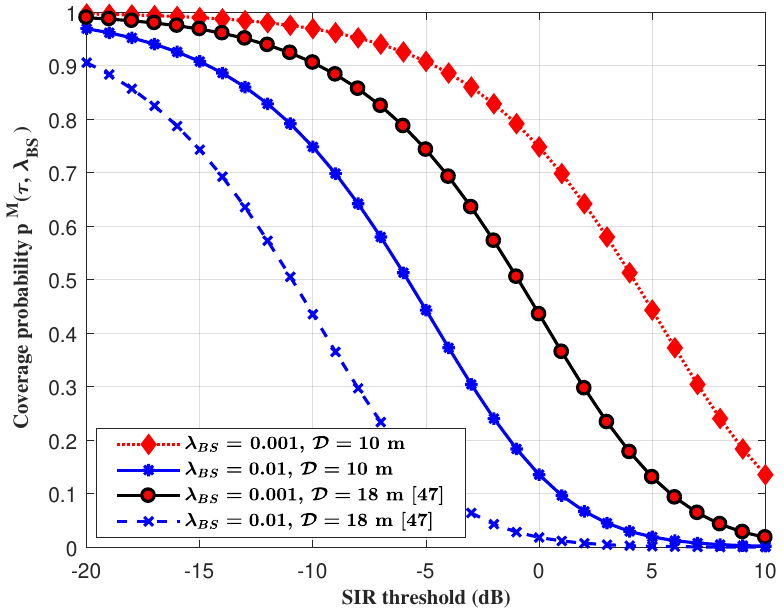}
\caption{Coverage probability with respect to the SIR threshold. }
\label{fig3}
\end{figure}

 Fig. \ref{fig3} depicts the coverage probability with respect to the SIR threshold $\tau $ considering different BS densities ${{\lambda }_{BS}}$ and critical distance ${\mathcal{D}}$. The results of Fig. \ref{fig3} show that the coverage probability decreases as the SIR threshold increases. When the SIR threshold and BS density are fixed, the coverage probability decreases with the increase of the critical distance. When the SIR threshold and critical distance are fixed, the coverage probability decreases as BS density increases considering the fixed number of cooperating BSs.

\section{GROUP-CELL HANDOVER SKIPPING SCHEME}
To reduce the handover rate in the group cell scenarios, we propose a GCHO skipping (GCHO-S) scheme that the UE dynamically skips the forthcoming GCHO processes along the corresponding trajectory. The rationale behind the handover skipping\footnote{Handover skipping approach allows the UE to skip multiple associating BSs, i.e., sacrifice the best connectivity along their trajectory to reduce the handover rate and minimize the handover delay \cite{b42}.} approach is to reduce the frequent GCHO, hence harvesting user-centric cooperation benefits in UDNs.

\subsection{GCHO-S Scheme}
\begin{figure}[!h]
\centering
\includegraphics[width=0.45\textwidth]{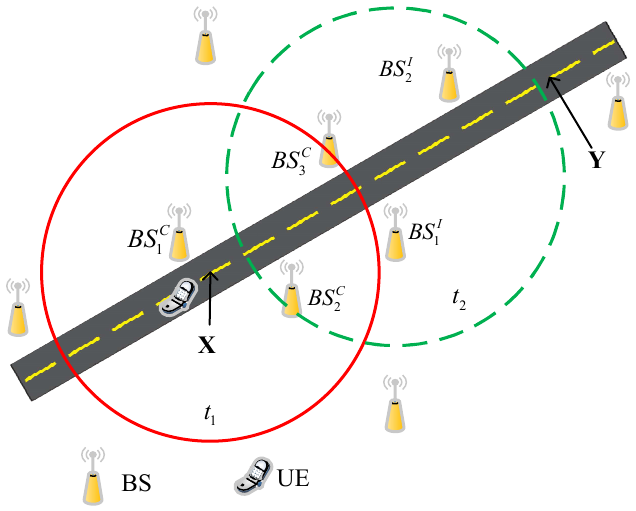}
\caption{GCHO skipping demonstration. The red circle shows the group-cell at ${{t}_{1}}$ and the green dotted circle indicates the group-cell at ${{t}_{2}}$ after the GCHO skipping when the UE ${{U}_{v}}$ moves from point X to point Y. }
\label{fig4}
\end{figure}

As previously discussed, the distances among the UE ${{U}_{v}}$ and cooperating BSs are denoted as ${{r}_{1,}}{{r}_{2,}}{{r}_{3}}.....{{r}_{M}}$. Without loss of generality, the number of cooperating BSs is configured as $M=3$, in this section. The proposed analysis can be adopted for $M>3$. Moreover, the initial order of the distance among the UE ${{U}_{v}}$ and cooperating BSs is assumed as ${{r}_{1}}<{{r}_{2}}<{{r}_{3}}$. When the UE ${{U}_{v}}$ crosses the group-cell boundaries, the distance relationship is changed to ${{r}_{3}}<{{r}_{2}}<{{r}_{1}}$. $BS_{1}^{I}$ and $BS_{2}^{I}$ represent the first and second nearest BSs, which are outside the group-cell region, respectively. $r_{1}^{I}$ represents the distance between the UE ${{U}_{v}}$ and the BS $BS_{1}^{I}$, while $r_{2}^{I}$ denotes the distance between the UE ${{U}_{v}}$ and the BS $BS_{2}^{I}$.For the skipping to occur, the distances are assumed as $r_{1}^{I}<{{r}_{M}}$ and $r_{2}^{I}\le 2{r_{1}^{I}}$ when the UE is at the group-cell boundary. In this case, the GCHO skipping is triggered by BS $BS_{1}^{I}$, i.e., the UE ${{U}_{v}}$ skips the association with the BS $BS_{1}^{I}$ to reconfigure the group-cell and execute the GCHO with BS $BS_{2}^{I}$ to avoid frequent GCHO and ping-pong GCHO as well. During the GCHO skipping process, the UE ${{U}_{v}}$ is not associated with the nearest BS $BS_{1}^{I}$ to form the new group-cell, although the UE ${{U}_{v}}$ is located within the skipped BS $BS_{1}^{I}$ vicinity. We term this as the skip-phase, where BSs $BS_{2}^{I}$, $BS_{3}^{C}$ and $BS_{2}^{C}$ jointly serve the UE ${{U}_{v}}$, as Fig. \ref{fig4} depicts. Algorithm 1 illustrates the proposed GCHO-S scheme.\\
\begin{algorithm}[htp]
\caption{Proposed GCHO-S Scheme }
\renewcommand{\algorithmicrequire}{\textbf{Input:}}
\renewcommand{\algorithmicensure}{\textbf{Output:}}
\label{alg1}
\begin{algorithmic}[1]
    \REQUIRE $\mathbb{C}$, ${{r}_{g}}$, where$\,\,g=1....M$, $\Psi $, ${{\Theta }_{{{r}_{M}}}}$, ${{\mathfrak{T}}_{v}}$, $r_{1}^{I}$, $r_{2}^{I}$;

    \ENSURE Updated group-cell set $\mathbb{C}$;
    \STATE Initialization;
    \STATE Set skipDone = false;\\
    \STATE \textbf{While} $\left( {{\mathfrak{T}}_{v}}\bigcap {{\Theta }_{{{r}_{M}}}} \right)$ \textbf{do}
    \STATE \hspace{10pt} \textbf{if} $\left( \begin{aligned}
  & r_{1}^{I}<{{r}_{M}}\,\,\And \And \,\,r_{2}^{I}\le 2r_{1}^{I} \\
 & \,\And \And \,\,\text{skipDone !}=\text{ true} \\
\end{aligned} \right)$ \textbf{then}
    \STATE \hspace{20pt}Perform GCHO skipping ();   $\Rightarrow$ skip $BS_{1}^{I}$
    \STATE \hspace{20pt}Set skipDone = true;
    \STATE \hspace{30pt}\textbf{else} Perform GCHO ();
    \STATE \hspace{30pt} Set skipDone = false;
    \STATE \hspace{30pt}\textbf{end else}
    \STATE \hspace{20pt}\textbf{end if}
   \STATE \textbf{end while}
  \STATE Result: update $\mathbb{C}$;
\end{algorithmic}
\end{algorithm}

For the proposed GCHO-S scheme (Fig. \ref{fig4}, algorithm 1), the handover rate is obtained as
\begin{equation}\label{eqn20}
  H_{M}^{*}=\frac{\Psi \sqrt{{{\lambda }_{BS}}}}{\sqrt{\pi }\sqrt{M}}.
\end{equation}
\subsection{Group-cell Size Optimization}
We consider the weighted sum to evaluate the tradeoff between the handover rate and cooperation signaling overhead in the group-cell. We assume that all handovers are successful in the GCHO process.  Let ${{S}_{1}}$ and ${{S}_{2}}$ denote the cost of one handoff and the cost of one CSI feedback message, respectively. Moreover, we assume that all handovers have the same cost value ${{S}_{1}}$, and all cooperation signaling overhead have the same cost value ${{S}_{2}}$. Consequently, the overall cost of the UE ${{U}_{v}}$  being served by $M$ BSs considering the GCHO and GCHO-S schemes are given in (\ref{eqn21a}) and (\ref{eqn21b}), respectively.
\begin{subequations}
\begin{equation}\label{eqn21a}
{{C}_{GCHO}}={{S}_{1}}{{H}_{M}}+{{S}_{2}}\Upsilon _{K},
\end{equation}
\begin{equation}\label{eqn21b}
 {{C}_{GCHO-S}}={{S}_{1}}H_{M}^{*}+{{S}_{2}}\Upsilon _{K}.
\end{equation}
\end{subequations}
The overall cost ${{C}_{GCHO}}$ and ${{C}_{GCHO-S}}$ are the unitless quantities and proportional to the percentage of time wasted due to handover and cooperation signaling processing, assuming no useful data is transmitted during all the processes. In our analysis, ${{S}_{1}}={{t}_{H}}$. By substituting (\ref{eqn11}) and (\ref{eqn3}) into (\ref{eqn21a}), we have,
\begin{equation}\label{eqn22}
  {{C}_{GCHO}}={{S}_{1}}\frac{2\Psi \sqrt{{{\lambda }_{BS}}}}{\sqrt{\pi }\sqrt{M}}+{{S}_{2}}M\frac{\mu }{T}.
\end{equation}

Denote ${{M}^*}$ as the optimal group-cell size, i.e., the optimal $M$ that minimizes  ${{C}_{GCHO}}$ in the GCHO scheme. By setting the first order derivative of  ${{C}_{GCHO}}$ to zero, the ${M}^*$ can be expressed as
\begin{equation}\label{eqn23}
 M^*={{\left( \frac{{{S}_{1}}^{2}{{\Psi }^{2}}{{T}^{2}}{{\lambda }_{BS}}}{\pi {{\mu }^{2}}{{S}_{2}}^{2}} \right)}^{\frac{1}{3}}}.
\end{equation}
By following a similar approach as in (\ref {eqn23}), the optimal group-cell size for the GCHO-S skipping scheme is given as
\begin{equation}\label{eqn24}
   {M}^{**}={{\left( \frac{{{S}_{1}}^{2}{{\Psi }^{2}}{{T}^{2}}{{\lambda }_{BS}}}{4\pi {{\mu }^{2}}{{S}_{2}}^{2}} \right)}^{\frac{1}{3}}}.
\end{equation}
\section{SIMULATION RESULTS AND DISCUSSION}
Numerical analysis is carried out to evaluate the effectiveness of our proposed handover schemes in user-centric cooperative transmissions. The single BS association handover and the fixed-region (FR) user-centric cooperative network topology handover \cite{b31} approaches are regarded as benchmarks and are referred to as “Traditional-HO” and “FR-HO”, respectively. Default simulation parameters are as follows: the path loss exponents are ${{\eta }_{1}}= 2$ and ${{\eta }_{2}}= 4$, the critical distance ${\mathcal{D}}$= 10 m, ${{t}_{H}}$= 0.3 s \cite{b32}, $\mu$= 1, ${{S}_{2}}$= 0.01${{T}}$ and ${{T}}$= 5 ms \cite{b39}.

\begin{figure}[!h]
\centering
\includegraphics[width=0.45\textwidth]{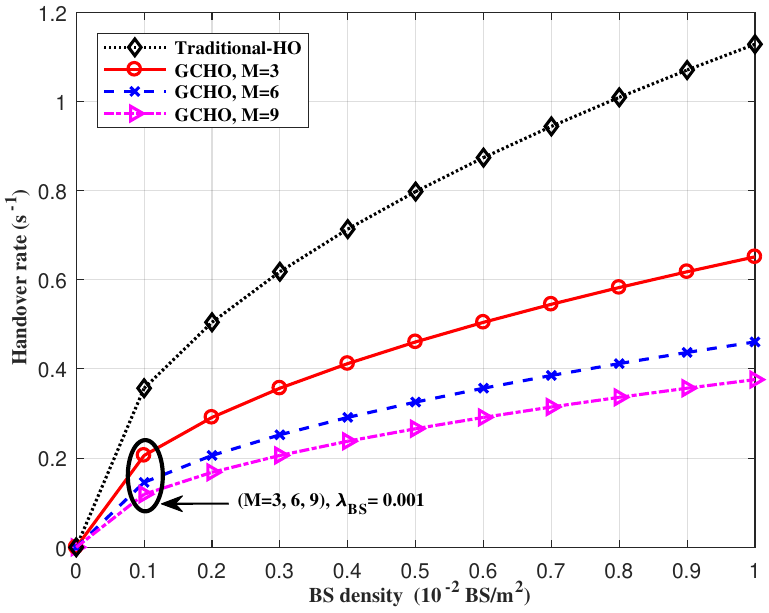}
\caption{Handover rate with respect to the BS density for $\Psi$ = 10 m/s.}
\label{fig5}
\end{figure}
Fig. \ref{fig5} shows the handover rate with respect to the BS density considering different group-cell sizes. Fig. \ref{fig5} indicates that higher BS density induces more handover rate when the group-cell size is fixed. When the BS density is fixed, the handover rate decreases as the group-cell size increases. Compared with the traditional handover approach, the shrinkage of handover rate  of the GCHO approach is 42.3\%, 59.2\% and 66.7\% at ${M}$=3, 6 and 9, respectively, when the BS density is ${{\lambda }_{BS}}$ = 0.001 BS/${m}^{2}$.

\begin{figure}[!h]
\centering
\includegraphics[width=0.45\textwidth]{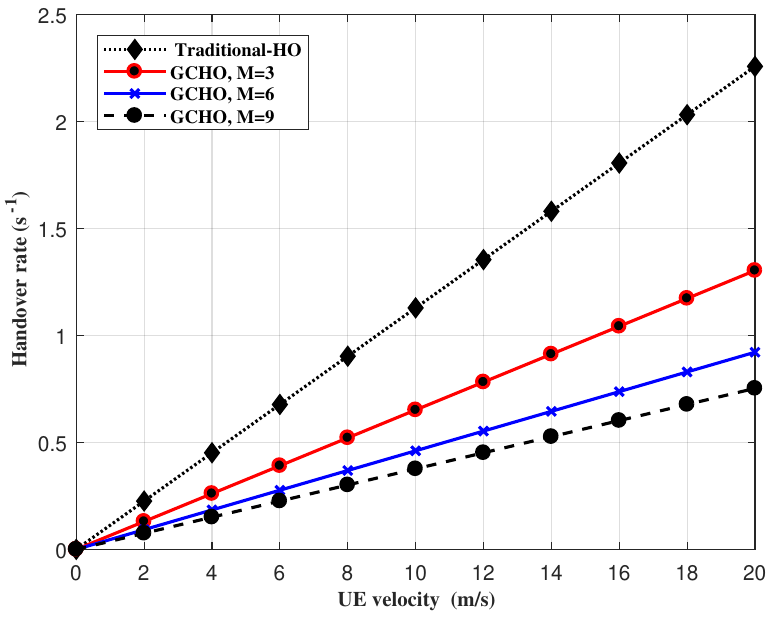}
\caption{Handover rate with respect to the UE velocity for ${{\lambda }_{BS}}$ = 0.01 BS/${m}^{2}$. }
\label{fig6}
\end{figure}
Fig. \ref{fig6} analyzes the impact of the group-cell size on the handover rate considering different UE velocities. The handover rate increases with the increase of UE velocity. When the UE velocity is fixed, the handover rate decreases with the rise of the group-cell size.

\begin{figure}[!h]
\centering
\includegraphics[width=0.45\textwidth]{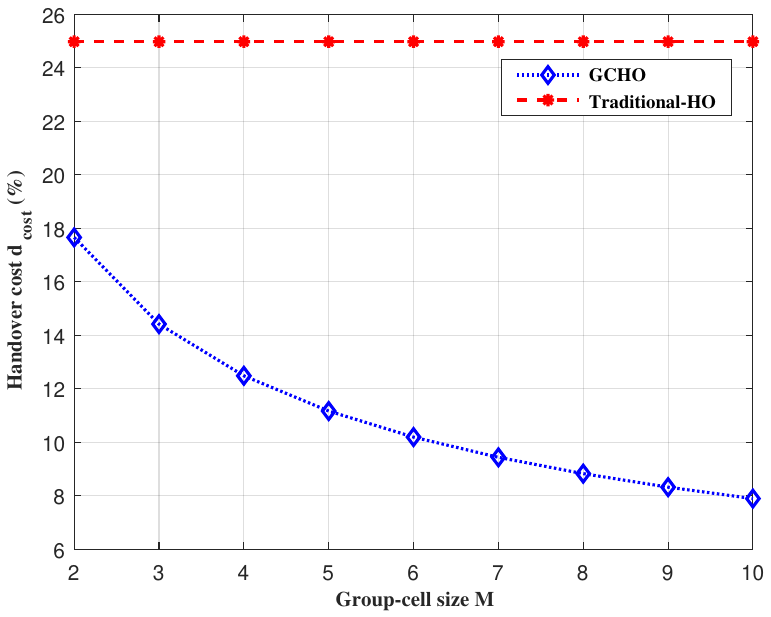}
\caption{Handover cost with respect to the group-cell size for ${{\lambda }_{BS}}$ = 0.01 BS/${m}^{2}$ and $\Psi$ = 10 m/s.}
\label{fig7}
\end{figure}
Fig. \ref{fig7} shows the handover cost with respect to the group-cell size. It can be seen that the handover cost reduces with the increase of the group-cell size for the proposed GCHO scheme. Considering (\ref{eqn18}), the handover cost is directly proportional to the handover signaling delay. Thus, the GCHO scheme is more efficient in reducing the delay resulting from the handover processes in contrast to the traditional handover approach.

\begin{figure}[!h]
\centering
\includegraphics[width=0.45\textwidth]{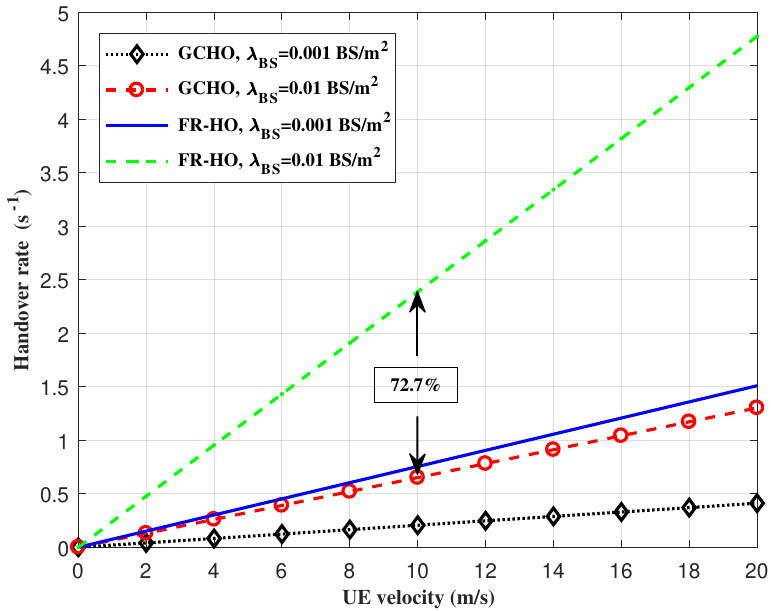}
\caption{Handover rate with respect to the UE velocity considering different handover schemes with $M$ = 3.}
\label{fig8}
\end{figure}
Fig. \ref{fig8} depicts the handover rate with respect to the UE velocity considering the GCHO and FR-HO schemes under various BS densities. Fig. \ref{fig8} demonstrates that as UE velocity and BS density increase, the handover rate increases for GCHO and FR-HO schemes. When the UE velocity is fixed, the GCHO scheme can reduce the handover rate by 72.7\% compared with the FR-HO approach.

\begin{figure}[!h]
\centering
\includegraphics[width=0.45\textwidth]{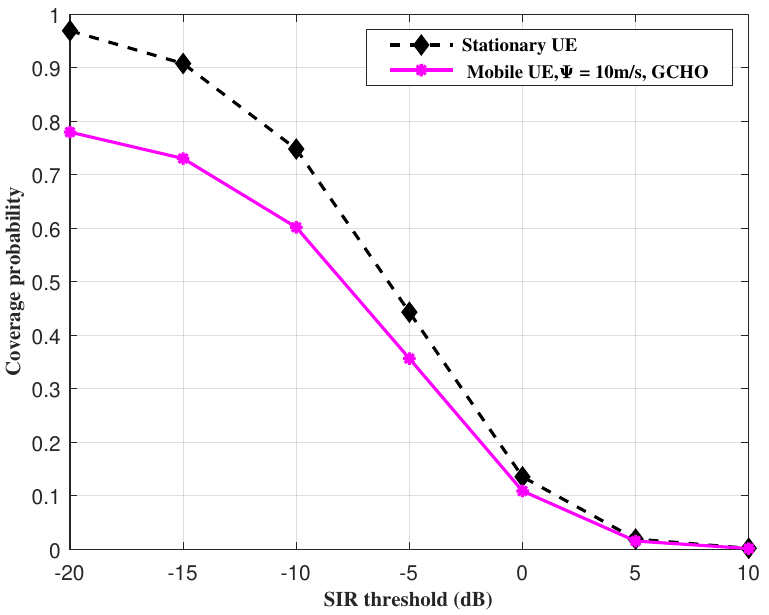}
\caption{Coverage probability with respect to the SIR threshold under the dual-slope path loss model for $M$ = 3 and ${{\lambda }_{BS}}$ = 0.01 BS/${m}^{2}$. }
\label{fig9}
\end{figure}
Fig. \ref{fig9} shows the coverage probability with respect to the SIR threshold considering the stationary and mobile UE. When the UE velocity is fixed, the coverage probability of the mobile UE deteriorates  faster than the coverage of the stationary UE as the SIR threshold increases, reflecting the fact that the SIR of the mobile UE is severely affected by the UE velocity.

\begin{figure}[!h]
\centering
\includegraphics[width=0.45\textwidth]{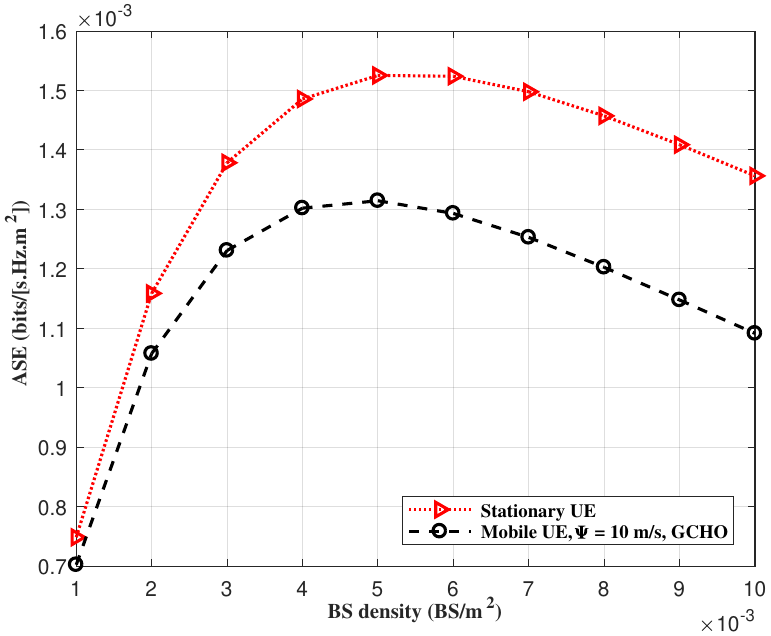}
\caption{ASE with respect to BS density under the dual-slope path loss model for $M$ = 3 and $\tau$= 0 dB. }
\label{fig10}
\end{figure}
Fig. \ref{fig10} investigates the impact of the handover on ASE with varying BS density. As indicated by Fig. \ref{fig10}, the ASE of the mobile UE is less than the ASE of the stationary UE. The percentage of the ASE reduction for the mobile UE increases as the BS density increases. Compared with the stationary UE, the ASE of the mobile UE decreases by 8.3\% and 21.43\% when the BS density is 0.002 BS/${m}^{2}$ and 0.01 BS/${m}^{2}$, respectively. Moreover, there exists an optimal BS density value to maximize the ASE of the stationary and mobile UE.

\begin{figure}[!h]
\centering
\includegraphics[width=0.45\textwidth]{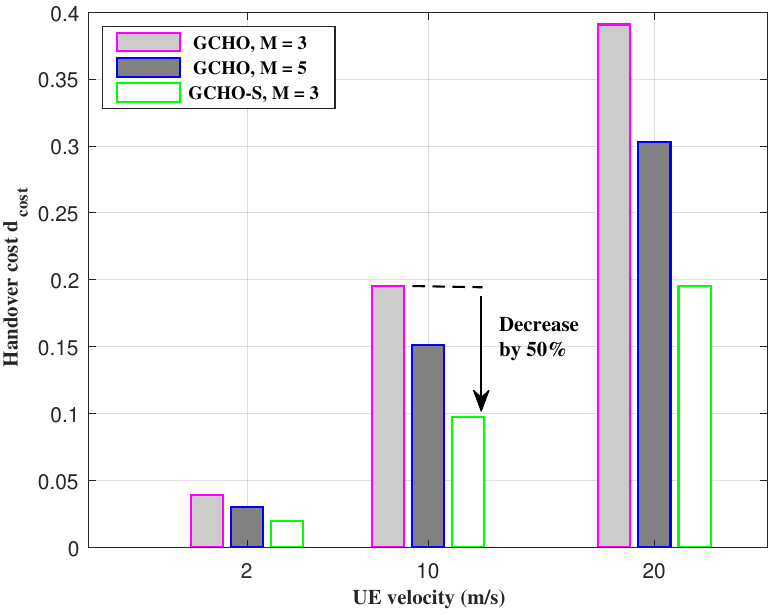}
\caption{Handover cost with respect to UE velocity for ${{\lambda }_{BS}}$ = 0.01 BS/${m}^{2}$.}
\label{fig11}
\end{figure}
Fig. \ref{fig11} evaluates the handover cost between the GCHO and GCHO-S schemes with respect to the UE velocity. Fig. \ref{fig11} portrays that the GCHO-S scheme reduces the handover cost as velocity increases in contrast with the best-connected GCHO scheme. Under the same group-cell size, the GCHO-S scheme can sufficiently diminish the handover cost up to 50\% against the GCHO scheme for various UE velocities.

\begin{figure}[!h]
\centering
\includegraphics[width=0.45\textwidth]{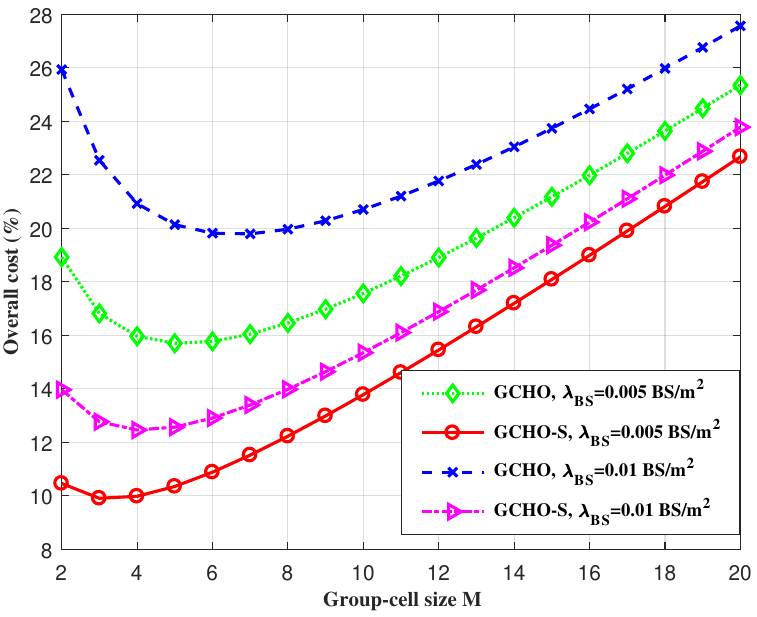}
\caption{Overall cost with respect to the group-cell size for $\Psi$= 10 m/s. }
\label{fig12}
\end{figure}
Fig. \ref{fig12} illustrates the overall cost with respect to the group-cell size considering various BS densities. As can be seen, the overall cost minimizes at an optimal group-cell size considering different BS densities in GCHO and GCHO-S schemes. When the group-cell size is fixed, the GCHO-S scheme yields a lower overall cost compared with the GCHO scheme. Moreover, on average, the cluster size of $M$= 3 can provide the lower overall cost when the BS density is 0.005 BS/${m}^{2}$. This result is suggested for the practical implementation of the GCHO and GCHO-S schemes to ensure the satisfactory operation of the SDN controller.

\begin{figure}[!h]
\centering
\includegraphics[width=0.45\textwidth]{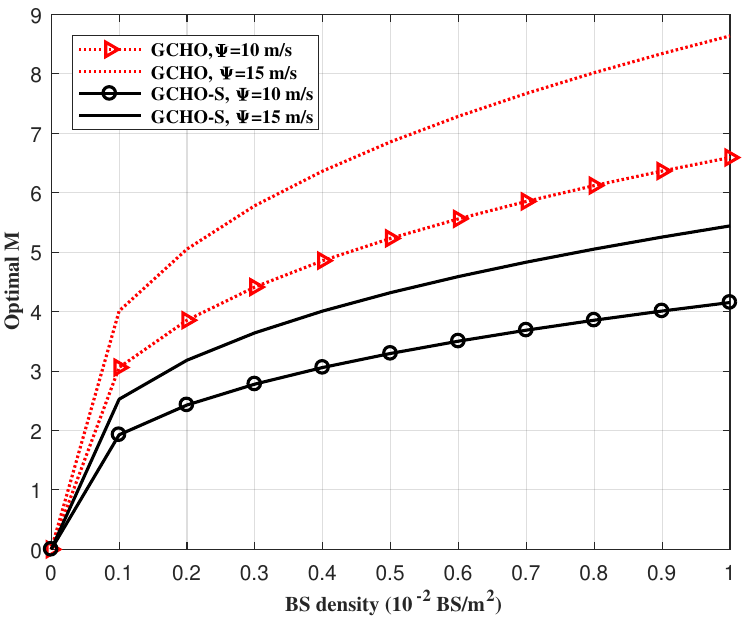}
\caption{Optimal group-cell size with respect to the BS density.}
\label{fig13}
\end{figure}
Fig. \ref{fig13} depicts the optimal group-cell size with respect to the BS density under different UE velocities. When the UE velocity is fixed, the optimal group-cell size increases as BS density increases in GCHO and GCHO-S schemes. Also, the results show that the optimal group-cell size is an ascending function of the UE velocity when BS density is fixed. However, the optimal group-cell size of the GCHO-S scheme is smaller than that of the GCHO scheme in all settings.
\section{CONCLUSION}
In this paper, a new GCHO scheme under the user-centric cooperative transmissions is proposed to improve the handover performance in UDNs. Based on stochastic geometry tools, the BSs locations distributions are modeled as homogenous PPP to realize the BSs cooperation region commonly encountered in practice. Then, the closed-form expression of the handover rate for an arbitrary UE movement is derived. Furthermore, the GCHO-S scheme where the UE skips connection with the forthcoming best group-cell is proposed to diminish the handover cost in user-centric cooperative transmissions deployments. And notably, an optimal group-cell size is necessary to achieve a desirable system adopting the GCHO and GCHO-S schemes. Thus, a weighted sum of the handover rate and cooperating signaling overhead is derived to optimize the group-cell size. On the other hand, numerical results show that the proposed GCHO scheme adequately reduces the handover rate in contrast to the Traditional HO and FR-HO approaches by 42.3\% and 72.7\%, respectively. Also, the proposed GCHO-S scheme may suppress the handover cost up to 50\% compared with the GCHO scheme, considering the same group-cell size. In addition, it is revealed that the optimal group-cell size is sensitive to variations in both the UE velocity and BS density. We further conclude that the GCHO-S scheme has demonstrated a good trade-off between the handover rate and cooperation signaling overhead due to the reduced handover cost. The cluster size of $M$ = 3 is recommended for the satisfactory operation of the SDN controller. This result can lay a foundation in subsequent investigations of the handover performance improvement, particularly for the delay-sensitive applications in which a small group-cell size is a significant criterion.  Moreover, the GCHO algorithm complexity and response time reduction as group-cell size increases may be an interesting topic for future work.
\appendices
\section{Proof of Theorem 1}
With the assumption that the BSs are randomly distributed and the UE is always located at the origin, the area intensity ${{\Omega }_{2}}({{\Theta }_{{{\delta }_{M}}}})$ of ${{\Theta }_{{{\delta }_{M}}}}$ is the probability that the origin ${{Q}_{0}}$ is in extended group-cell boundary ${{\Theta }_{{{\delta }_{M}}}}$, i.e., $\mathbb{P}({{Q}_{0}}\in {{\Theta }_{{{\delta }_{M}}}})$.
\begin{equation}\label{eqn25}
  \begin{aligned}
  & {{\Omega }_{2}}({{\Theta }_{{{\delta }_{M}}}})=\mathbb{P}({{Q}_{0}}\in {{\Theta }_{{{\delta }_{M}}}}) \\
 & \,\,\,\,\,\,\,\,\,\,\,\,\,\,\,\,\,\,\,=\frac{2{{\delta }_{M}}}{{{r}_{M}}}+O({{\delta }_{M}}^{2})\,\,. \\
\end{aligned}
\end{equation}
\section{Proof of Theorem 2}
The coverage probability in (\ref{eqn14}) is redefined as (\ref{eqn26}) , where (a) follows from the complementary cumulative distribution function (CCDF) of the Gamma distribution and the assumption that the UE is positioned at the edge of the cooperating BSs. The distance between the UE and the cooperating BS is approximated to be equal for the cell-edge UE in the group-cell and is denoted as $R$ (b) follows from the Laplace transform of interference, i.e., the random variable $I(M)$ , where $\mathcal{L}_{I(M)}^{n}\left( s \right)$ stands for $n-th$ derivative of ${{\mathcal{L}}_{I(M)}}\left( s \right)$ and $\mathcal{L}_{I(M)}^{n}\left( s \right)={{\left( -1 \right)}^{n}}{{\mathbb{E}}_{I(M)}}\left( I{{(M)}^{n}}{{e}^{-sI(M)}} \right)={{\left( -1 \right)}^{n}}\frac{{{d}^{n}}}{d{{s}^{n}}}{{\mathcal{L}}_{I(M)}}\left( s \right)$.
\begin{equation} \label{eqn26}
  \begin{aligned}
  & p^{M}(\tau ,{{\lambda }_{BS}})\triangleq \mathbb{P}(SIR(M)>\tau ) \\
 & \,\,\,\,\,\,\,\,\,\,\,\,\,\,\,\,\,\,\,\,\,\,=\mathbb{P}\left( \frac{\sum\limits_{g=1}^{M}{{{p}_{g}}r_{g}^{-{{\eta }_{1}}}{{h}_{g}}}}{I(M)}>\tau  \right)\, \\
 & \,\,\,\,\,\,\,\,\,\,\,\,\,\,\,\,\,\,\,\,={{\mathbb{E}}_{{{r}_{g}}}}\left( \mathbb{P}\left( \sum\limits_{g=1}^{M}{r_{g}^{-{{\eta }_{1}}}{{h}_{g}}}>\tau I\left( M \right) \right) \right) \\
 & \,\,\,\,\,\,\,\,\,\,\,\,\,\,\,\,\,\,\,\overset{(a)}{\mathop{=}}\,{{\mathbb{E}}_{R}}\left( \sum\limits_{n=0}^{M-1}{\frac{{{\left( \frac{\tau I\left( M \right)}{{{R}^{-{{\eta }_{1}}}}} \right)}^{n}}}{n!}{{e}^{-\frac{\tau I\left( M \right)}{{{R}^{-{{\eta }_{1}}}}}}}} \right) \\
 & \,\,\,\,\,\,\,\,\,\,\,\,\,\,\,\,\,\,={{\mathbb{E}}_{R}}{{\mathbb{E}}_{I(M)}}\left( \sum\limits_{n=0}^{M-1}{\frac{{{\left( \frac{\tau }{{{R}^{-{{\eta }_{1}}}}} \right)}^{n}}}{n!}I{{(M)}^{n}}{{e}^{-\frac{\tau I\left( M \right)}{{{R}^{-{{\eta }_{1}}}}}}}} \right) \\
 & \,\,\,\,\,\,\,\,\,\,\,\,\,\,\,\,\,\overset{(b)}{\mathop{=}}\,{{\mathbb{E}}_{R}}\left( \sum\limits_{n=0}^{M-1}{\frac{{{\left( -\frac{\tau }{{{R}^{-{{\eta }_{1}}}}} \right)}^{n}}}{n!}\mathcal{L}_{I(M)}^{n}\left( \frac{\tau }{{{R}^{-{{\eta }_{1}}}}} \right)} \right) \\
 & \,\,\,\,\,\,\,\,\,\,\,\,\,\,\,\,={{\mathbb{E}}_{R}}\left( \sum\limits_{n=0}^{M-1}{\frac{{{\left( -s \right)}^{n}}}{n!}\mathcal{L}_{I(M)}^{n}\left( s \right)} \right), \\
\end{aligned}
\end{equation}

The Laplace function ${{\mathcal{L}}_{I(M)}}\left( s \right)$ for $s=\frac{\tau }{{{R}^{-{{\eta }_{1}}}}}$ is further derived as
\begin{equation} \label{eqn27}
\begin{aligned}
  & {{\mathcal{L}}_{I(M)}}\left( s \right)={{\mathbb{E}}_{I(M)}}\left[ {{e}^{-sI(M)}} \right]\\
  & \,\,\,\,\,\,\,\,\,\,\,\,={{\mathbb{E}}_{\Phi_{BS} ,{{h}_{g}}}}\left[ \exp \left( -s\sum\limits_{g\in \Phi_{BS} >M}{\Lambda r_{g}^{-{{\eta }_{2}}}{{h}_{g}}} \right) \right] \\
 & \,\,\,\,\,\,\,\,\,\,\,\,={{\mathbb{E}}_{\Phi_{BS} ,{{h}_{g}}}}\left[ \prod\limits_{g\in \Phi_{BS} >M}{\exp \left( -s\Lambda r_{g}^{-{{\eta }_{2}}}{{h}_{g}} \right)} \right] \\
 & \,\,\,\,\,\,\,\,\,\,\,\,={{\mathbb{E}}_{\Phi_{BS} }}\left[ \prod\limits_{g\in \Phi_{BS} >M}{{{\mathbb{E}}_{{{h}_{g}}}}\left[ \exp \left( -s\Lambda r_{g}^{-{{\eta }_{2}}}{{h}_{g}} \right) \right]} \right], \\
\end{aligned}
\end{equation}
\begin{equation}\label{eqn28}
\begin{aligned}
  & {{\mathcal{L}}_{I(M)}}\left( s \right)\,\overset{(c)}{\mathop{=}}\,{{\mathbb{E}}_{\Phi_{BS} }}\left[ \prod\limits_{g\in \Phi_{BS} >M}{\frac{1}{1+s\Lambda r_{g}^{-{{\eta }_{2}}}}} \right] \\
 & \,\,\,\,\,\,\,\,\,\,\,\,\,\,\,\,\,\,\,\,\,\,\overset{(d)}{\mathop{=}}\,\exp \left( -2\pi {{\lambda }_{BS}}\int\limits_{{{r}_{g}}}^{\infty }{\left( 1-\frac{1}{1+s\Lambda r_{g}^{-{{\eta }_{2}}}} \right)wdw} \right), \\
\end{aligned}
\end{equation}
where (c) follows from ${{h}_{g}}\sim \exp (1)$ is exponential distributed (d) follows from probability generating functional (PGFL) \cite{b45} of the PPP, which states for some function $f(y)$ that $\mathbb{E}\left[ \prod\nolimits_{y\in \Phi_{BS} }{f(y)} \right]=\exp \left( -\lambda_{BS} \int_{{{\mathbb{R}}^{2}}}{\left( 1-f(y \right)dy} \right)$.
By employing the change of variables $p={{w}^{2}}$ and $dp=2wdw$ then,
\begin{equation} \label{eqn29}
    {{\mathcal{L}}_{I(M)}}\left( s \right)=\exp \left( -\pi {{\lambda }_{BS}}\int\limits_{r_{g}}^{\infty }{\left( 1-\frac{1}{1+s\Lambda {{p}^{-\tfrac{{{\eta }_{2}}}{2}}}} \right)dp} \right).
\end{equation}
By using the change of variables  $\,u={{\left( s\Lambda  \right)}^{-\tfrac{2}{{{\eta }_{2}}}}}p$ and $dp={{\left( s\Lambda  \right)}^{\tfrac{2}{{{\eta }_{2}}}}}du$, then
\begin{equation}\label{eqn30}
    {{\mathcal{L}}_{I(M)}}\left( s \right)=\exp \left( -\pi {{\lambda }_{BS}}{{\left( s\Lambda  \right)}^{\frac{2}{{{\eta }_{2}}}}}\int\limits_{\vartheta }^{\infty }{\left( \frac{1}{1+{{u}^{\tfrac{{{\eta }_{2}}}{2}}}} \right)du} \right),
\end{equation}
where $\vartheta =\frac{{{R}^{2}}}{{{\left( s\Lambda  \right)}^{\frac{2}{{{\eta }_{2}}}}}}$.
By substituting $\Lambda =\mathcal D^{^{\left( {{\eta }_{2}}-{{\eta }_{1}} \right)}}$and $s=\frac{\tau }{{{R}^{-{{\eta }_{1}}}}}$ into (\ref{eqn30}), then
\begin{equation}\label{eqn31}
  {{\mathcal{L}}_{I(M)}}\left( s \right)=\exp \left( \begin{aligned}
  & -\pi {{\lambda }_{BS}}{{\left( \tau \mathcal D^{{{\eta }_{2}}-{{\eta }_{1}}}{{R}^{{{\eta }_{1}}}} \right)}^{\frac{2}{{{\eta }_{2}}}}} \\
 & \times \int\limits_{\vartheta }^{\infty }{\left( \frac{1}{1+{{u}^{\tfrac{{{\eta }_{2}}}{2}}}} \right)du} \\
\end{aligned} \right).
\end{equation}
Next, let
\begin{equation}\label{eqn32}
  {{a}_{0}}={{\mathcal{L}}_{I(M)}}\left( s \right)=\exp (-\pi {{\lambda }_{BS}}{{k}_{0}}),
\end{equation}
 where,
\begin{equation}\label{eqn33}
    {{k}_{0}}={{\left( \tau \mathcal D^{{{\eta }_{2}}-{{\eta }_{1}}}{{R}^{{{\eta }_{1}}}} \right)}^{\frac{2}{{{\eta }_{2}}}}}\int\limits_{\vartheta }^{\infty }{\left( \frac{1}{1+{{u}^{\tfrac{{{\eta }_{2}}}{2}}}} \right)du}.
\end{equation}
Then, (\ref{eqn26}) can be rewritten as
\begin{equation} \label{eqn34}
\begin{aligned}
p^{M}\left(\tau, \lambda_{BS}\right) &=\mathbb{E}_{R}\left(\sum_{n=0}^{M-1} a_{n}\right) \\
&=\mathbb{E}_{R}\left(a_{0}+\sum_{n=1}^{M-1} a_{n}\right).
\end{aligned}
\end{equation}
where
\begin{equation}\label{eqn35}
    {{a}_{n}}=\frac{{{s}^{n}}}{n!}{{\left( -1 \right)}^{n}}\mathcal{L}_{I(M)}^{n}\left( s \right).
\end{equation}

The $n-th$ derivative of the Laplace transform in (\ref{eqn35}) can be expressed in a recursive form. Eventually, the coverage probability can be represented by a lower triangular Toeplitz matrix. Following \cite{b57}, the $n-th$ derivative can be written as
\begin{equation} \label{eqn36}
\begin{aligned}
  & \mathcal{L}_{I(M)}^{n}\left( s \right)=\pi {{\lambda }_{BS}}\sum\limits_{i=0}^{n-1}{\left( \begin{aligned}
  & n-1 \\
 & \,\,\,\,i \\
\end{aligned} \right)}{{\left( -1 \right)}^{n-i}}\left( n-i \right)!{{\left( s\Lambda  \right)}^{\frac{2}{{{\eta }_{2}}}-n+i}} \\
 & \ \ \ \times \int\limits_{\vartheta }^{\infty }{\frac{{{\left( {{u}^{-\tfrac{{{\eta }_{2}}}{2}}} \right)}^{n-i}}}{\,\,\,{{\left( 1+{{u}^{-\tfrac{{{\eta }_{2}}}{2}}} \right)}^{n-i+1}}}du}\times \mathcal{L}_{I(M)}^{i}\left( s \right).
\end{aligned}
\end{equation}
Furthermore, ${{a}_{n}}$ in (\ref{eqn35}) is derived by the formula (\ref{eqn37}) by substituting (\ref{eqn36}) into (\ref{eqn35}).\\
\begin{figure*}
\begin{equation}\label{eqn37}
  \begin{aligned}
  & {{a}_{n}}=\pi {{\lambda }_{BS}}\sum\limits_{i=0}^{n-1}{\left( \begin{aligned}
  & n-1 \\
 & \,\,\,\,i \\
\end{aligned} \right)}\frac{\left( n-i \right)!}{n!}{{\left( -1 \right)}^{n}}{{\left( -1 \right)}^{n-i}}{{s}^{n}}{{\left( s\Lambda  \right)}^{\frac{2}{{{\eta }_{2}}}-n+i}}\times \underbrace{\int\limits_{\vartheta }^{\infty }{\frac{{{\left( {{u}^{-\tfrac{{{\eta }_{2}}}{2}}} \right)}^{n-i}}}{\,\,\,{{\left( 1+{{u}^{-\tfrac{{{\eta }_{2}}}{2}}} \right)}^{n-i+1}}}du}}_{\varepsilon (s)}\times \mathcal{L}_{I(M)}^{i}\left( s \right) \\
 & {{a}_{n}}=\pi {{\lambda }_{BS}}\sum\limits_{i=0}^{n-1}{\left( \begin{aligned}
  & n-1 \\
 & \,\,\,\,i \\
\end{aligned} \right)}\frac{\left( n-i \right)!}{n!}{{s}^{n}}{{\left( s\Lambda  \right)}^{\frac{2}{{{\eta }_{2}}}-n+i}}\times \underbrace{\int\limits_{\vartheta }^{\infty }{\frac{{{\left( {{u}^{-\tfrac{{{\eta }_{2}}}{2}}} \right)}^{n-i}}}{\,\,\,{{\left( 1+{{u}^{-\tfrac{{{\eta }_{2}}}{2}}} \right)}^{n-i+1}}}du}}_{\varepsilon (s)}\times {{\left( -1 \right)}^{i}}\mathcal{L}_{I(M)}^{i}\left( s \right) \\
 & \,\,\,\,\,=\pi {{\lambda }_{BS}}\sum\limits_{i=0}^{n-1}{\left( \begin{aligned}
  & n-1 \\
 & \,\,\,\,i \\
\end{aligned} \right)}\frac{\left( n-i \right)!}{n!}{{s}^{\frac{2}{{{\eta }_{2}}}}}{{\left( \Lambda  \right)}^{\frac{2}{{{\eta }_{2}}}-n+i}}\varepsilon (s)\times {{s}^{i}}{{\left( -1 \right)}^{i}}\mathcal{L}_{I(M)}^{i}\left( s \right) \\
 & \,\,\,\,=\pi {{\lambda }_{BS}}\sum\limits_{i=0}^{n-1}{\left( \begin{aligned}
  & n-1 \\
 & \,\,\,\,i \\
\end{aligned} \right)}\frac{\left( n-i \right)!i!}{n!}{{s}^{\frac{2}{{{\eta }_{2}}}}}{{\left( \Lambda  \right)}^{\frac{2}{{{\eta }_{2}}}-n+i}}\varepsilon (s)\times {{a}_{i}} \\
 & \,\,\,\overset{(e)}{\mathop{=}}\,\pi {{\lambda }_{BS}}{{\left( s\Lambda  \right)}^{\frac{2}{{{\eta }_{2}}}}}\sum\limits_{i=0}^{n-1}{\frac{n-i}{n}}{{\Lambda }^{-n+i}}\varepsilon (s)\times {{a}_{i}} \\
\end{aligned},
\end{equation}
\hrule
\end{figure*}
where step (e) follows from a combination property. Next, by substituting the values of $s$ and $\Lambda $ into (\ref{eqn37}), we obtain (\ref{eqn38})
\begin{equation} \label{eqn38}
\begin{aligned}
  & {{a}_{n}}=\pi {{\lambda }_{BS}}{{\left( \frac{\tau \mathcal D^{{{\eta }_{2}}-{{\eta }_{1}}}}{{{R}^{-{{\eta }_{1}}}}} \right)}^{\frac{2}{{{\eta }_{2}}}}}\sum\limits_{i=0}^{n-1}{\frac{n-i}{n}}\mathcal D^{\left( {{\eta }_{2}}-{{\eta }_{1}} \right)\left( -n+i \right)} \\
 & \times \int\limits_{\vartheta }^{\infty }{\frac{{{\left( {{u}^{-\tfrac{{{\eta }_{2}}}{2}}} \right)}^{n-i}}}{\,\,\,{{\left( 1+{{u}^{-\tfrac{{{\eta }_{2}}}{2}}} \right)}^{n-i+1}}}du}\times {{a}_{i}},
\end{aligned}
\end{equation}
\begin{equation}\label{eqn39}
\begin{aligned}
& {{a}_{n}}=\,\pi {{\lambda }_{BS}}{{\left( \tau \mathcal D^{{{\eta }_{2}}-{{\eta }_{1}}}{{R}^{{{\eta }_{1}}}} \right)}^{\frac{2}{{{\eta }_{2}}}}}\sum\limits_{i=0}^{n-1}{\frac{n-i}{n}}\mathcal D^{\left( {{\eta }_{2}}-{{\eta }_{1}} \right)\left( -n+i \right)}{{k}_{n-i}}{{a}_{i}},
\end{aligned}
\end{equation}
 where ${{k}_{i}}=\int\limits_{\vartheta }^{\infty }{\frac{1}{\,\,\,{{\left( 1+{{u}^{\tfrac{{{\eta }_{2}}}{2}}} \right)}^{i}}\left( 1+{{u}^{-\tfrac{{{\eta }_{2}}}{2}}} \right)}du}.$   \,\,\,\,\,     for  $i\ge 1$\\
Furthermore, ${{a}_{n}}$ can be solved in explicit expression using linear algebra manipulation. After iterating, ${{a}_{n}}$ can be rewritten as
\begin{equation} \label{eqn40}
    {{a}_{n}}={{b}_{0}}{{a}_{0}}\mathcal D^{-\left( {{\eta }_{2}}-{{\eta }_{1}} \right)}{{k}_{n}}+\sum\limits_{i=1}^{n-1}{{{b}_{0}}}F_{n}^{i}{{a}_{i}}
\end{equation}
where ${{b}_{0}}=\pi {{\lambda }_{BS}}{{\left( \tau \mathcal D^{{{\eta }_{2}}-{{\eta }_{1}}}{{R}^{{{\eta }_{1}}}} \right)}^{\frac{2}{{{\eta }_{2}}}}}$ and ${{F}_{n}}$ is $n\times n$ lower triangular Toeplitz matrix as $F_{n}^{i}=0$ for $i\ge n$, ${{F}_{n}}$ is given as (\ref{eqn41}), \\

Next, we express ${{A}_{M-1}}=\,\sum\limits_{n=1}^{M-1}{{{a}_{n}}}$, then the coverage probability is
\begin{figure*}[!t]
 \begin{equation}\label{eqn41}
    {{F}_{n}}=\left[ \begin{matrix}
   0 & {} & {} & {} & {} & {}  \\
   \frac{1}{2}\mathcal D^{-\left( {{\eta }_{2}}-{{\eta }_{1}} \right)}{{k}_{1}} & 0 & {} & {} & {} & {}  \\
   \frac{2}{3}\mathcal D^{-2\left( {{\eta }_{2}}-{{\eta }_{1}} \right)}{{k}_{2}}\, & \frac{1}{3}\mathcal D^{-\left( {{\eta }_{2}}-{{\eta }_{1}} \right)}{{k}_{1}} & 0 & {} & {} & {}  \\
   \frac{3}{4}\mathcal D^{-3\left( {{\eta }_{2}}-{{\eta }_{1}} \right)}{{k}_{3}}\,\, & \frac{2}{4}\mathcal D^{-2\left( {{\eta }_{2}}-{{\eta }_{1}} \right)}{{k}_{2}} & \frac{1}{4}\mathcal D^{-\left( {{\eta }_{2}}-{{\eta }_{1}} \right)}{{k}_{1}} & 0 & {} & {}  \\
   \vdots  & {} & {} & {} & 0 & {}  \\
   \frac{M-1}{M}\mathcal D^{\left( {{\eta }_{2}}-{{\eta }_{1}} \right)(-M+1)}{{k}_{M-1}} & \frac{M-2}{M}\mathcal D^{\left( {{\eta }_{2}}-{{\eta }_{1}} \right)(-M+2)}{{k}_{M-2}} & \cdots  & \cdots  & \frac{1}{M}\mathcal D^{-\left( {{\eta }_{2}}-{{\eta }_{1}} \right)}{{k}_{1}} & 0  \\
\end{matrix} \right].
 \end{equation}
 \hrule
\end{figure*}
\begin{equation}\label{eqn42}
    \begin{aligned}
  & p^{M}(\tau ,{{\lambda }_{BS}})={{\mathbb{E}}_{R}}\left( {{a}_{0}}+{{A}_{M-1}} \right) \\
 & \,\,\,\,\,\,\,\,\,\,\,\,\,\,\,\,\,\,\,\,\,\,\,\,=\int\limits_{0}^{\infty }{\left( {{a}_{0}}+{{A}_{M-1}} \right){{f}_{(R)}}}(R)dR. \\
\end{aligned}
\end{equation}
Without loss of generality, in this setup, the group-cell size is three. Then, the distance distribution of $R$ in the group-cell is given by \cite{b58},
\begin{equation}\label{eqn43}
    {{f}_{R}}(R)=2{{(\pi {{\lambda }_{BS}})}^{2}}{{R}^{3}}{{e}^{-\pi {{\lambda }_{BS}}{{R}^{2}}}}.
\end{equation}
The coverage probability is further expressed as (\ref{eqn16}), which completes the proof.


\ifCLASSOPTIONcaptionsoff
  \newpage
\fi

\begin{IEEEbiography}[{\includegraphics[width=1in,height=1.25in,clip,keepaspectratio]{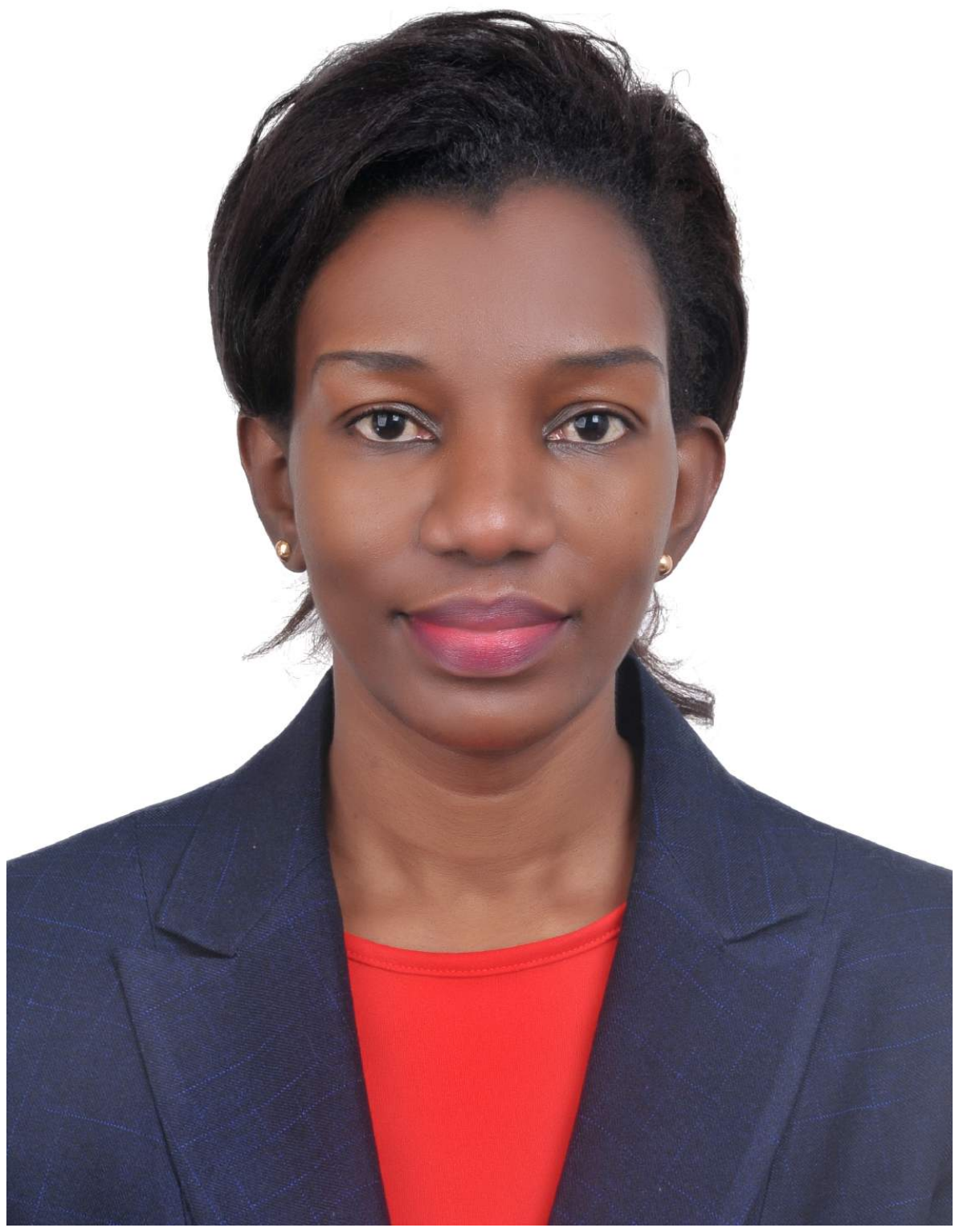}}]{Nyaura Kibinda }
(Student Member, IEEE) received the BSc. degree in telecommunication engineering from the University of Dar es Salaam (UDSM), Tanzania, in 2010 and MSc. degree in telecommunication engineering from the University of Dodoma (UDOM), Tanzania, in 2014. She is currently pursuing the Ph.D. degree in information and communication engineering at Huazhong University of Science and Technology (HUST), China. Her research interests include wireless communication networks, network modeling and handover management.
\end{IEEEbiography}

\vspace{-5 mm}
\begin{IEEEbiography}[{\includegraphics[width=1in,height=1.25in,clip,keepaspectratio]{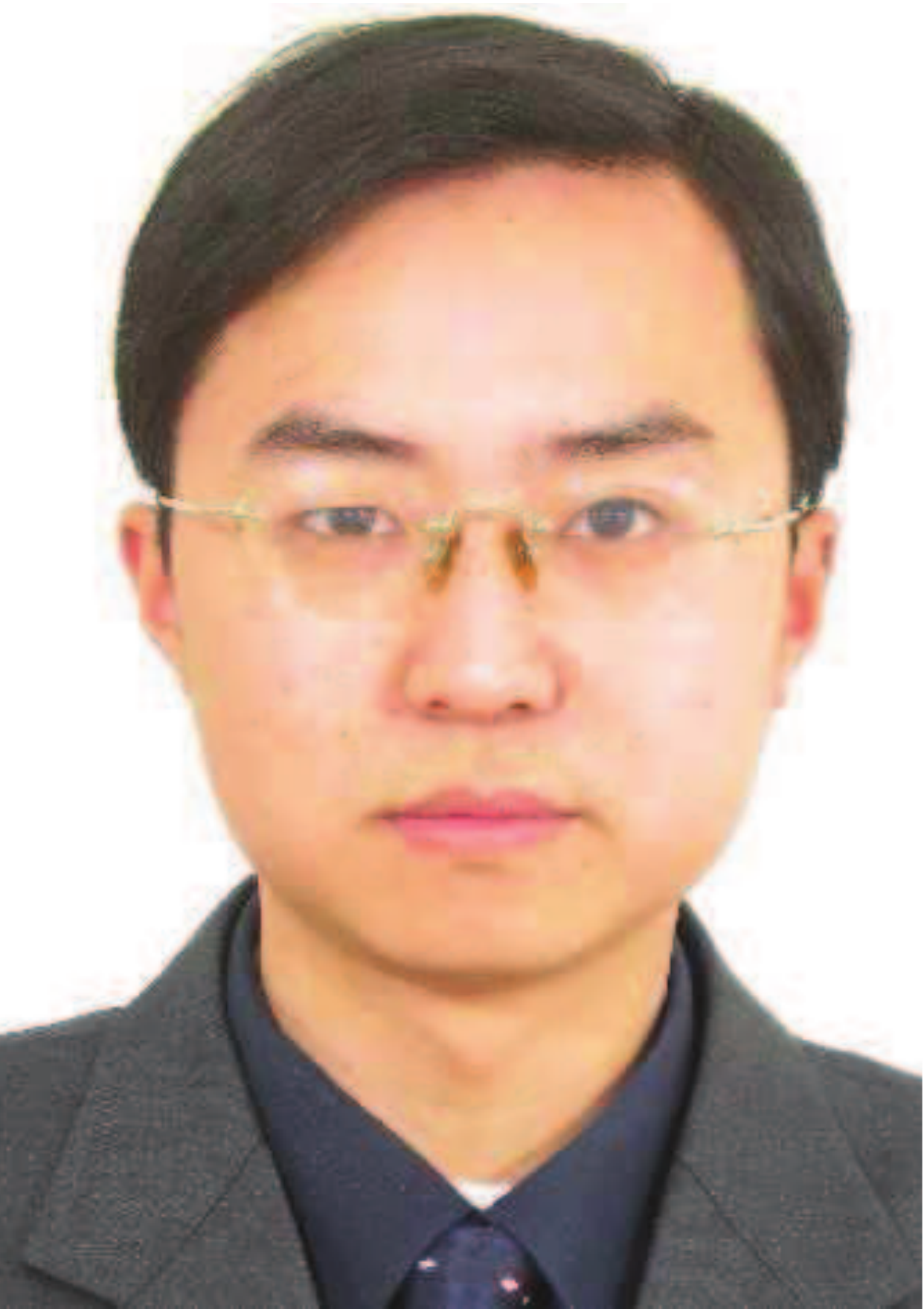}}]{Xiaohu Ge}
(Senior Member, IEEE) received the Ph.D. degree in communication and information engineering from the Huazhong University of Science and Technology (HUST), China, in 2003. He has been working with HUST, since November 2005. Prior to that, he worked as a Researcher at Ajou University,South Korea, and the Politecnico Di Torino, Italy, from January 2004 to October 2005. He is currently a
Full Professor with the School of Electronic Information and Communications, HUST. He is also an Adjunct Professor with the Faculty of Engineering and Information Technology, University of Technology Sydney (UTS), Australia. He has published about 200 articles in refereed journals and conference proceedings. He has been granted about 25 patents in China. His research interests are in the area of mobile communications, traffic modeling in wireless networks, green communications, and interference modeling in wireless communications. He services as an IEEE Distinguished Lecturer and an Associate Editor for the IEEE WIRELESS COMMUNICATIONS, IEEE ACCESS, IEEE TRANSACTIONS ON VEHICULAR TECHNOLOGY, etc.
\end{IEEEbiography}


\end{document}